\documentclass[twocolumn,floatfix,showpacs,rmp]{revtex4}
\usepackage{graphicx}
\usepackage{amsmath}
\usepackage{bm}
\begin{document}
\title{Andreev reflection and Klein tunneling in graphene}
\author{C. W. J. Beenakker}
\affiliation{Instituut-Lorentz, Universiteit Leiden, P.O. Box 9506, 2300 RA Leiden, The Netherlands}
\date{October 2007}
\begin{abstract}
This is a colloquium-style introduction to two electronic processes in a carbon monolayer (graphene), each having an analogue in relativistic quantum mechanics. Both processes couple electron-like and hole-like states, through the action of either a superconducting pair potential or an electrostatic potential. The first process, Andreev reflection, is the electron-to-hole conversion at the interface with a superconductor. The second process, Klein tunneling, is the tunneling through a \textit{p-n} junction. The absence of backscattering, characteristic of massless Dirac fermions, implies that both processes happen with unit efficiency at normal incidence. Away from normal incidence, retro-reflection in the first process corresponds to negative refraction in the second process. In the quantum Hall effect, both Andreev reflection and Klein tunneling induce the same dependence of the two-terminal conductance plateau on the valley isospin of the carriers. Existing and proposed experiments on Josephson junctions and bipolar junctions in graphene are discussed from a unified perspective.
\medskip\\
{\tt "Colloquium" for Reviews of Modern Physics.}
\end{abstract}
\maketitle
\tableofcontents

\section{\label{intro} Introduction}

\begin{figure}[tb]
\includegraphics[width=0.9\linewidth]{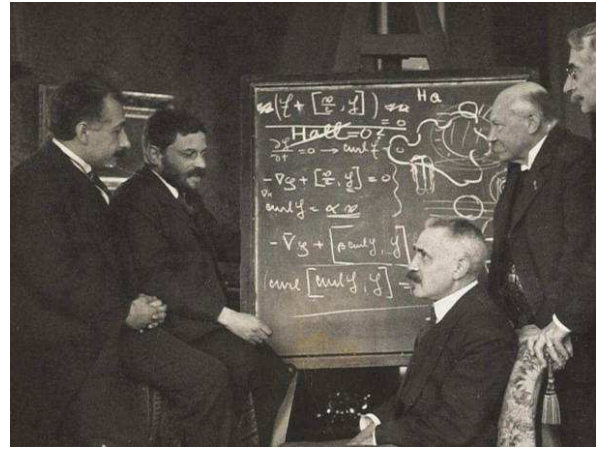}
\caption{\label{EEKO}
Albert Einstein, Paul Ehrenfest, Paul Langevin, Heike Kamerlingh Onnes, and Pierre Weiss at a workshop in Leiden (October, 1920). The blackboard discussion, on the Hall effect in superconductors, has been reconstructed by \textcite{Sau07}. See also \textcite{Del06} for the historical context of this meeting.
}
\end{figure}

In October 1920 the inventor of special relativity, Albert Einstein, traveled to Leiden to meet with the discoverer of superconductivity, Heike Kamerlingh Onnes. A photograph of a blackboard (Fig.\ \ref{EEKO}) records one of their discussions. The two physicists had much to discuss, but they would have found little common ground in the two topics closest to their hearts, since superconductivity is essentially a nonrelativistic phenomenon.

Relativistic contributions to the superconducting pair potential, studied by \textcite{Cap95,Cap99}, are a small correction of order $(v_{F}/c)^{2}$ [Fermi velocity over speed of light squared]. Fully relativistic phenomena such as particle-to-antiparticle conversion by a superconductor have remained pure fiction. Some of this fiction is now becoming science in a material first isolated a few years ago by Andre Geim and his group at Manchester University \cite{Nov04}.

The material, called graphene, is a mono-atomic layer of carbon atoms arranged on a honeycomb lattice. Upon doping, electrons and holes move through the layer with a velocity $v=10^{6}\,{\rm m/s}$ which is only a small fraction of the speed of light. And yet, this velocity is energy independent --- as if the electrons and holes were massless particles and antiparticles moving at the speed of light. As demonstrated in transport measurements by \textcite{Nov05} and \textcite{Zha05}, and in spectroscopic measurements by \textcite{Zho06} and \textcite{Bos07}, the electronic properties of graphene are described by an equation (the Dirac equation) of relativistic quantum mechanics, even though the microscopic Hamiltonian of the carbon atoms is nonrelativistic. While graphene itself is not superconducting, it acquires superconducting properties by proximity to a superconductor. We therefore have the unique possibility to bridge the gap between relativity and superconductivity in a real material.

\begin{figure}[tb]
\includegraphics[width=0.8\linewidth]{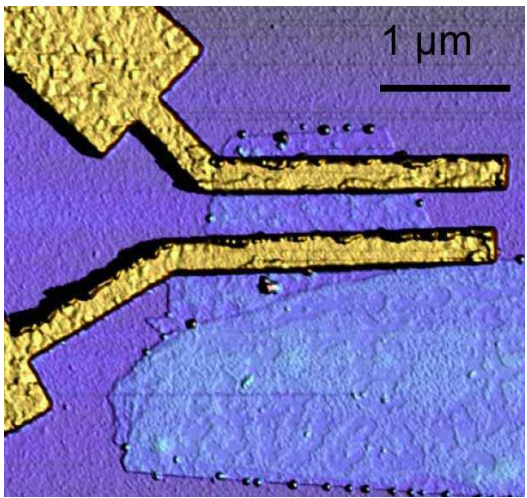}
\caption{\label{SNSmonolayer}
Atomic force microscope image (false color) of a carbon monolayer covered by two superconducting Al electrodes. \cite{Hee07}
}
\end{figure}

For example, Fig.\ \ref{SNSmonolayer} shows two superconducting electrodes on top of a carbon monolayer. The supercurrent measured through this device by \textcite{Hee07} is carried by massless electrons and holes, converted into each other by the superconducting pair potential. This conversion process, known as Andreev reflection \cite{And64}, is described by a superconducting variant of the Dirac equation \cite{Bee06}.

In this Colloquium we review the unusual physics of Andreev reflection in graphene. For a broader perspective, we compare and contrast this coupling of electrons and holes by a superconducting pair potential with the coupling of electrons and holes by an electrostatic potential. The latter phenomenon is called Klein tunneling \cite{Che06,Kat06} with reference to relativistic quantum mechanics, where it represents the tunneling of a particle into the Dirac sea of antiparticles \cite{Kle29}. Klein tunneling in graphene is the tunneling of an electron from the conduction band into hole states from the valence band (which plays the role of the Dirac sea).

The two phenomena, Andreev reflection and Klein tunneling, are introduced in Secs.\ \ref{Andreev} and \ref{Klein}, respectively, and then compared in Sec.\ \ref{Analogies}. But first we summarize, in Sec.\ \ref{basics}, the special properties of graphene that govern these two phenomena. More comprehensive reviews of graphene have been written by \textcite{Cas06,Kat07a,Kat07b,Gei07,Gus07,Cas07}.

\section{\label{basics}Basic physics of graphene}
\subsection{\label{Diracequation}Dirac equation}

\begin{figure}[tb]
\includegraphics[width=0.8\linewidth]{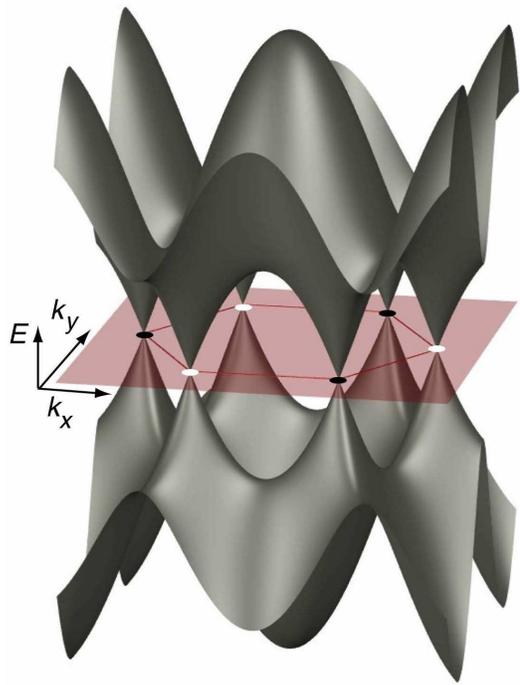}
\caption{\label{GrapheneBand}
Band structure $E(k_{x},k_{y})$ of a carbon monolayer. The hexagonal first Brillouin zone is indicated. The conduction band ($E>0$) and the valence band ($E<0$) form conically shaped valleys that touch at the six corners of the Brillouin zone (called conical points, or Dirac points, or $K$-points). The three corners marked by a white dot are connected by reciprocal lattice vectors, so they are equivalent. Likewise, the three corners marked by a black dot are equivalent. In undoped graphene the Fermi level passes through the Dirac points. (Illustration by C. Jozsa and B. J. van Wees.)
}
\end{figure}

\begin{figure}[tb]
\includegraphics[width=0.8\linewidth]{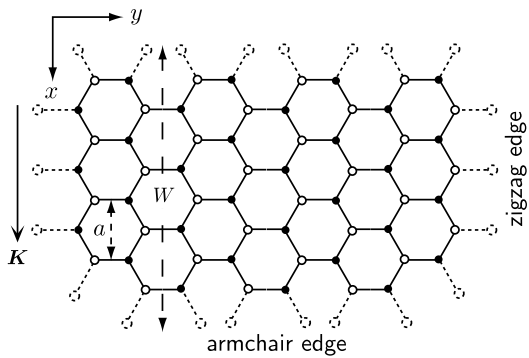}
\caption{\label{monolayer}
Honeycomb lattice of a carbon monolayer. The unit cell contains two atoms, labeled $A$ and $B$, each of which generates a triangular sublattice (open and closed circles). The lattice constant $a$ is $\sqrt{3}$ times larger than the carbon-carbon separation of $0.142\,{\rm nm}$. The reciprocal lattice vector $\bm{K}$ has length $4\pi/3a$. The edge of the lattice may have the armchair configuration (containing an equal number of atoms from each sublattice), or the zigzag configuration (containing atoms from one sublattice only). Dashed circles and bonds indicate missing atoms and dangling bonds, respectively. The separation $W$ of opposite edges is measured from one row of missing atoms to the opposite row, as indicated in the figure.
}
\end{figure}

The unusual band structure of a single layer of graphite, shown in Fig.\ \ref{GrapheneBand}, has been known for 60 years \cite{Wal47}. Near each corner of the hexagonal first Brillouin zone the energy $E$ has a conical dependence on the two-dimensional wave vector $\bm{k}=(k_{x},k_{y})$. Denoting by $\delta\bm{k}=\bm{k}-\bm{K}$ the displacement from the corner at wave vector $\bm{K}$, one has for $|\delta\bm{k}|a\ll 1$ the dispersion relation
\begin{equation}
|E|=\hbar v|\delta\bm{k}|.\label{Evk}
\end{equation}
The velocity $v\equiv\frac{1}{2}\sqrt{3}\tau a/\hbar\approx 10^{6}\,{\rm m/s}$ is proportional to the lattice constant $a=0.246\,{\rm nm}$ and to the nearest-neighbor hopping energy $\tau\approx 3\,{\rm eV}$ on the honeycomb lattice of carbon atoms (shown in Fig.\ \ref{monolayer}).

The linear dispersion relation \eqref{Evk} implies an energy-independent group velocity $v_{\rm group}\equiv |\partial E/\hbar\partial\bm{k}|=v$ of low-energy excitations ($|E|\ll\tau$). These electron excitations (filled states in the conduction band) or hole excitations (empty states in the valence band) therefore have zero effective mass. \textcite{Sem84}, and \textcite{DiV84}  noticed that --- even though $v\ll c$ --- such massless excitations are governed by a wave equation, the Dirac equation, of relativistic quantum mechanics:
\begin{equation}
-i\hbar v\begin{pmatrix}
0&\partial_{x}-i\partial_{y}\\
\partial_{x}+i\partial_{y}&0
\end{pmatrix}\begin{pmatrix}
\Psi_{A}\\
\Psi_{B}
\end{pmatrix}
=E \begin{pmatrix}
\Psi_{A}\\
\Psi_{B}
\end{pmatrix}.\label{Diraconevalley}
\end{equation}
[The derivation of this equation for a carbon monolayer goes back to \textcite{McC56}.]

The two components $\Psi_{A}$ and $\Psi_{B}$ give the amplitude $\Psi_{A}(\bm{r})e^{i\bm{K}\cdot\bm{r}}$ and $\Psi_{B}(\bm{r})e^{i\bm{K}\cdot\bm{r}}$ of the wave function on the $A$ and $B$ sublattices of the honeycomb lattice (see Fig.\ \ref{monolayer}). The differential operator couples  $\Psi_{A}$ to $\Psi_{B}$ but not to itself, in view of the fact that nearest-neighbor hopping on the honeycomb lattice couples only $A$-sites with $B$-sites,\footnote{Next-nearest-neighbor hopping contributes second-order spatial derivatives, which are of higher order in $a|\delta\bm{k}|$ and may therefore be neglected in first approximation.} 
\begin{equation}
E\psi_{A}=\tau\sum_{\rm neighbors}\psi_{B},\;\
E\psi_{B}=\tau\sum_{\rm neighbors}\psi_{A}.
\label{tightbinding}
\end{equation}

In a more concise notation, Eq.\ \eqref{Diraconevalley} may be written as
\begin{equation}
v\bm{p}\cdot\bm{\sigma}\psi=E\psi,
\label{Diraconevalley2}
\end{equation}
with $\bm{p}=-i\hbar(\partial/\partial x,\partial/\partial y)$ the momentum operator in the $x$-$y$ plane and $\bm{\sigma}=(\sigma_{x},\sigma_{y},\sigma_{z})$ the vector of Pauli matrices acting on the spinor $\psi=(\Psi_{A},\Psi_{B})$. (For later use we define $\sigma_{0}$ as the $2\times 2$ unit matrix.) The spin degree of freedom described by the Pauli matrices $\sigma_{i}$ is called the ``pseudospin'', to distinguish it from the real electron spin.

This two-dimensional Dirac equation describes states with wave vector $\bm{k}$ in the valley centered at the corner of the Brillouin zone with wave vector $\bm{K}=(4\pi/3a)\bm{\hat{x}}$. The valley at the opposite corner at $-\bm{K}$ produces an independent set of states with amplitudes $\Psi'_{A}(\bm{r})e^{-i\bm{K}\cdot\bm{r}}$ and $\Psi'_{B}(\bm{r})e^{-i\bm{K}\cdot\bm{r}}$ on the $A$ and $B$ sublattices. The two components $\Psi'_{A}$ and $\Psi'_{B}$ satisfy the same Dirac equation \eqref{Diraconevalley2} with $p_{x}\rightarrow -p_{x}$. The spinor $\Psi=(\Psi_{A},\Psi_{B},-\Psi'_{B},\Psi'_{A})$ containing both valleys therefore satisfies the four-dimensional Dirac equation\footnote{
The valley-isotropic representation \eqref{Diractwovalley} of the four-dimensional Dirac equation (with two identical $2\times 2$ subblocks) is used to write boundary conditions in a compact form (see Sec.\ \ref{boundaries}). Other representations (with two unequal subblocks) are common in the literature as well, and one should be aware of this when comparing formulas from different papers.}
\begin{equation}
\begin{pmatrix}
v\bm{p}\cdot\bm{\sigma}&0\\
0&v\bm{p}\cdot\bm{\sigma}
\end{pmatrix}\Psi=E\Psi.\label{Diractwovalley}
\end{equation}
This differential equation represents the low-energy and long-wave length limit of the difference equation \eqref{tightbinding} in the tight-binding model of graphene.

For a compact notation, we make use of a second set of Pauli matrices $\bm{\tau}=(\tau_{x},\tau_{y},\tau_{z})$, with $\tau_{0}$ the $2\times 2$ unit matrix, acting on the valley degree of freedom (while $\bm{\sigma}$ and $\sigma_{0}$ act on the sublattice degree of freedom). Eq.\ \eqref{Diractwovalley} may then be written as
\begin{subequations}
\label{HDirac}
\begin{align}
&H(\bm{A})\Psi=E\Psi,\label{HDiraca}\\
&H(\bm{A})=v[(\bm{p}+e\bm{A})\cdot\bm{\sigma}]\otimes\tau_{0}+U\sigma_{0}\otimes\tau_{0},\label{HDiracb}
\end{align}
\end{subequations}
where for generality we have also included external electromagnetic fields (with scalar potential $U$ and vector potential $\bm{A}$). Electromagnetic fields do not couple the two valleys, provided that the fields vary smoothly on the scale of the lattice constant.

To conclude this subsection we briefly comment on the quantum-relativistic analogue of Eq.\ \eqref{Diractwovalley}, referring to \textcite{Gus07} for a more extensive discussion. In three dimensions, and with a change of sign of one of the two subblocks $v\bm{p}\cdot\bm{\sigma}$, Eq.\ \eqref{Diractwovalley} represents the Dirac (or Dirac-Weyl) equation of massless neutrinos, with $v$ the speed of light. The valley degree of freedom corresponds to the chirality of the neutrinos, which have left-handed or right-handed circular polarization (corresponding to the opposite sign of the two subblocks). In two dimensions the relative sign of the two subblocks can be changed by a unitary transformation, so the distinction between left- or right-handedness cannot be made. Electrons in graphene are called ``chiral'' because their direction of motion is tied to the direction of the pseudospin. Indeed, the current operator
\begin{equation}
\bm{j}=v\bm{\sigma}\otimes\tau_{0}\label{jdef}
\end{equation}
is proportional to the pseudospin operator $\bm{\sigma}$, so that an electron moving in the $x$ or $y$-direction has a pseudospin pointing in the $x$ or $y$-direction. But because the pseudospin is two-dimensional, there is no analogue of circular polarization and therefore there is no left- or right-handedness in graphene.

\subsection{\label{TRS} Time reversal symmetry}

The time reverse of the state $\Psi_{X}e^{i\bm{K}\cdot\bm{r}}+\Psi'_{X}e^{-i\bm{K}\cdot\bm{r}}$ on the $X=A,B$ sublattice is the complex conjugate $\Psi_{X}^{\ast}e^{-i\bm{K}\cdot\bm{r}}+{\Psi'}^{\ast}_{X}e^{i\bm{K}\cdot\bm{r}}$. This implies that the time reverse of the spinor $\Psi=(\Psi_{A},\Psi_{B},-\Psi'_{B},\Psi'_{A})$ is ${\cal T}\Psi=({\Psi'}_{A}^{\ast},{\Psi'}_{B}^{\ast},-\Psi_{B}^{\ast},\Psi_{A}^{\ast})$. The time reversal operator ${\cal T}$ therefore has the form
\begin{equation}
{\cal T}=\begin{pmatrix}
0&0&0&1\\
0&0&-1&0\\
0&-1&0&0\\
1&0&0&0
\end{pmatrix}{\cal C}=
-(\tau_{y}\otimes\sigma_{y}){\cal C},\label{Tdef}
\end{equation}
with ${\cal C}$ the operator of complex conjugation. Notice that the time reversal operation interchanges the valleys \cite{Suz02}.

The time reverse of the Dirac Hamiltonian \eqref{HDirac} is 
\begin{equation}
{\cal T}H(\bm{A}){\cal T}^{-1}=H(-\bm{A}).\label{THTdef}
\end{equation}
As it should be, time reversal symmetry is preserved in the absence of a magnetic field.

The Dirac Hamiltonian satisfies another anti-unitary symmetry, ${\cal S}H(\bm{A}){\cal S}^{-1}=H(-\bm{A})$, with 
\begin{equation}
{\cal S}=\begin{pmatrix}
0&1&0&0\\
-1&0&0&0\\
0&0&0&1\\
0&0&-1&0
\end{pmatrix}{\cal C}=
i(\tau_{0}\otimes\sigma_{y}){\cal C}.\label{Sdef}
\end{equation}
This operator ${\cal S}$ does not interchange the valleys, unlike ${\cal T}$, but like ${\cal T}$ it does invert the sign of $\bm{p}$ and $\bm{\sigma}$. The operator ${\cal S}$, therefore, acts like a time reversal operator in a single valley. The ${\cal S}$-symmetry of the Dirac Hamiltonian is called a {\em symplectic\/} symmetry, while the ${\cal T}$-symmetry is called an {\em orthogonal\/} symmetry.\footnote{
A symplectic symmetry operator is an anti-unitary operator which squares to $-1$, while an orthogonal symmetry operator is an anti-unitary operator which squares to $+1$. Both ${\cal T}$ and ${\cal S}$ are anti-unitary (product of a unitary operator and complex conjugation), but ${\cal T}^{2}=1$ while ${\cal S}^{2}=-1$.}

Because it is not the true time reversal symmetry operator on the honeycomb lattice, the symplectic symmetry can be broken even in the absence of a magnetic field \cite{Ber87}. Consider the following two perturbations $\delta H$ of the Dirac Hamiltonian:
\begin{itemize}
\item
A mass term $\delta H=\mu(\bm{r})\sigma_{z}\otimes\tau_{z}$, generated for example by a sublattice dependent potential in the substrate \cite{Zho07}.
\item
A valley-dependent vector potential, $\delta H=ev[{\cal A}(\bm{r})\cdot\bm{\sigma}]\otimes\tau_{z}$, produced by straining the monolayer \cite{Mor06a,Mor06b}.\footnote{A ripple of diameter $R$ and height $H$ corresponds to a fictitious magnetic field of order $|B|\simeq(\hbar/ea)H^{2}/R^{3}$, of opposite sign in the two valleys.}
\end{itemize}
In both cases, ${\cal T}\delta H{\cal T}^{-1}=\delta  H$, so time reversal symmetry is preserved, while ${\cal S}\delta H{\cal S}^{-1}=-\delta H$, so the symplectic symmetry is broken.

Whether it is the ${\cal T}$-symmetry or the ${\cal S}$-symmetry that governs a transport property depends on whether the scattering processes couple valleys or not. A smoothly varying electrostatic potential does not cause intervalley scattering, so it is the presence or absence of the symplectic symmetry ${\cal S}$ that matters in this case. [For example, breaking of ${\cal S}$ destroys the weak antilocalization effect, even if ${\cal T}$ is preserved \cite{Suz02,McC06,Ale06}.] Andreev reflection at a superconductor does couple the valleys \cite{Bee06}, so there it is the true time reversal symmetry ${\cal T}$ that matters. [For example, breaking of ${\cal T}$ suppresses the supercurrent while breaking of ${\cal S}$ does not \cite{Hee07}.]

\subsection{\label{boundaries} Boundary conditions}

The Dirac equation needs to be supplemented by a boundary condition of the form $\Psi=M\Psi$ at the edge of the graphene sheet \cite{McC04}. Since edges are typically abrupt on the atomic scale, the boundary condition couples the valleys. Ignoring a possible local magnetization, we may assume that $M$ commutes with ${\cal T}$ --- meaning that the boundary condition itself does not break time reversal symmetry. The boundary condition then has the form \cite{Akh07b}
\begin{equation}
\Psi=M\Psi,\;\;M = (\bm{\nu} \cdot \bm{\tau})\otimes (\bm{n}\cdot \bm{\sigma}), \label{boundarycondition}
\end{equation}
parameterized by a pair of three-dimensional unit vectors $\bm{\nu}$ and $\bm{n}$. The vector $\bm{n}$ is constrained by $\bm{n}\cdot\bm{n}_{B}=0$, to ensure that no current leaks out through the boundary (with normal $\bm{n}_{B}$, pointing outward).

We give three examples of boundaries \cite{Ber87,Bre06b}:
\begin{itemize}
\item
A zigzag edge has either $\Psi_{A}=\Psi'_{A}=0$ or $\Psi_{B}=\Psi'_{B}=0$, depending on whether the row of missing atoms at the edge is on the $A$ or $B$ sublattice (see Fig.\ \ref{monolayer}). The corresponding boundary condition matrix $M$ has $\bm{\nu}=\pm\bm{\hat{z}}$, $\bm{n}=\bm{\hat z}$. Because opposite zigzag edges lie on different sublattices, the angle $\Phi$ between the vectors $\bm{\nu}$ on opposite edges equals $\pi$ --- irrespective of the separation of the edges.
\item
An armchair edge has $\Psi_{X}e^{i\bm{K}\cdot\bm{r}}+\Psi'_{X}e^{-i\bm{K}\cdot\bm{r}}=0$ for $X=A,B$ --- so that the wave function vanishes on both sublattices. This requires $\bm{\nu}\cdot\bm{\hat{z}}=0$, $\bm{n}=\bm{\hat{z}}\times\bm{n}_{B}$. The angle $\Phi=|\bm{K}|W+\pi$ now depends on the separation $W$ (as defined in Fig.\ \ref{monolayer}): $\Phi=\pi$ if $2W/a$ is a multiple of three and $\Phi=\pm\pi/3$ otherwise.
\item
Confinement by an infinite mass has $\bm{\nu}=\bm{\hat{z}}$, $\bm{n}=\bm{\hat{z}}\times\bm{n}_{B}$.
\end{itemize}

\begin{figure}[tb]
\includegraphics[width=0.6\linewidth]{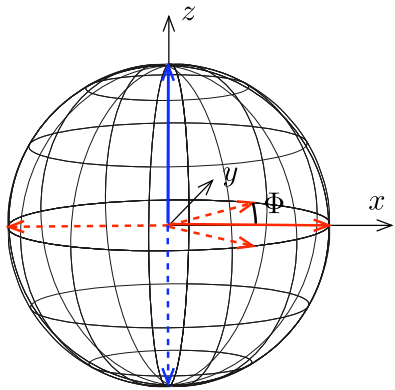}
\caption{\label{Blochsphere}
Location of the valley isospin $\bm{\nu}$ on the Bloch sphere for a zigzag edge (blue arrows) and for an armchair edge (red arrows). The solid and dashed arrows correspond to opposite edges.
}
\end{figure}

The two eigenstates $|+\bm{\nu}\rangle$ and $|-\bm{\nu}\rangle$ of $\bm{\nu}\cdot\bm{\tau}$ (defined by  $\bm{\nu}\cdot\bm{\tau}|\pm\bm{\nu}\rangle=\pm|\pm\bm{\nu}\rangle$) are states of definite valley polarization --- parallel or antiparallel to the unit vector $\bm{\nu}$. This vector is called the valley {\em isospin}, because it transforms under rotations in the same way as the real electron spin. It can be represented by a point on the Bloch sphere, see Fig.\ \ref{Blochsphere}. When $\bm{\nu}$ points in the $\bm{\hat z}$ direction, the polarization is such that the eigenstate lies entirely within one single valley. This is the case for the zigzag edge or for the infinite mass confinement. When $\bm{\nu}$ lies in the $x$-$y$ plane, the eigenstate is a coherent equal-weight superposition of the two valleys. This is the case for the armchair edge.

The direction of $\bm{\nu}$ in the boundary matrix $M$ plays a key role in a strong magnetic field, by selecting the valley polarization of edge states \cite{Akh07b}. Edge states in the lowest Landau level are valley polarized \cite{Bre06a,Aba06}, but the Hall conductance is insensitive to the direction $\bm{\nu}$ of the valley isospin.\footnote{
The Hall conductance $G_{H}=ge^{2}/h$ is determined by the degeneracy factor $g$ of the edge states. The celebrated ``half-integer'' Hall conductance $G_{H}=(n+1/2)\times 4e^{2}/h$ measured by \textcite{Nov05} and \textcite{Zha05} tells us that the lowest ($n=0$) Landau level has spin degeneracy but no valley degeneracy ($g=2$ rather than $g=4$). The direction of the valley polarization does not enter in $G_{H}$.} 
In Sec.\ \ref{QHE} we will see how Andreev reflection and Klein tunneling both provide a way to measure the valley isospin in the quantum Hall effect.

\begin{figure}[tb]
\includegraphics[width=0.8\linewidth]{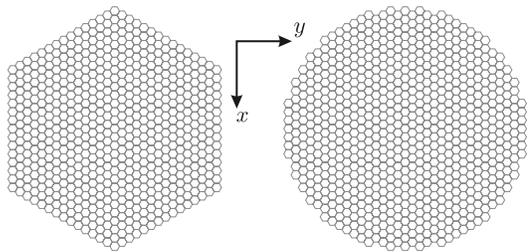}
\caption{\label{wow}
These two graphene flakes both have the same zigzag boundary condition: $\Psi=\pm\tau_{z}\otimes\sigma_{z}\Psi$. The sign switches between $+$ and $-$ at the armchair orientation (when the tangent to the boundary has an angle with the $y$-axis which is a multiple of $60^{\circ}$). \cite{Akh07c}
}
\end{figure}

We conclude this discussion of boundary conditions with the constraint imposed by electron-hole symmetry. In the absence of an electrostatic potential ($U=0$), the Dirac Hamiltonian \eqref{HDirac} anticommutes with $\tau_{z}\otimes\sigma_{z}$. In an unbounded system, this implies electron-hole symmetry of the spectrum. (If $\Psi$ is an eigenstate with eigenvalue $E$, then $\tau_{z}\otimes\sigma_{z}\Psi$ is an eigenstate with eigenvalue $-E$.) The electron-hole symmetry exists already at the level of the tight-binding model \eqref{tightbinding} [$E\mapsto -E$ if $\Psi_{B}\mapsto -\Psi_{B}$], so it is preserved by any boundary that is simply a termination of the lattice (zero edge potential).\footnote{
One mechanism that may produce an edge potential at a zigzag boundary (antiferromagnetic spin ordering) has been discussed in connection with graphene nanoribbons by \textcite{Son06}.
}
The requirement that the boundary matrix $M$ in Eq.\ \eqref{boundarycondition} commutes with $\tau_{z}\otimes\sigma_{z}$ (needed to preserve the electron-hole symmetry) restricts $M$ to either the zigzag form or the armchair form. As illustrated in Fig.\ \ref{wow}, the zigzag form is the rule while the armchair form is the exception \cite{Akh07c}.

\subsection{\label{pseudodif} Pseudo-diffusive dynamics}

Electrical conduction through a graphene sheet has unusual features when the Fermi level is at the Dirac point. Because of the vanishing density of states the transmission through a strip of undoped graphene (width $W$, length $L$ in the current direction) occurs entirely via evanescent (= exponentially decaying) modes. For a short and wide strip there is a large number $W/L\gg 1$ of evanescent modes with transmission probability of order unity. In a remarkable coincidence,\footnote{
We say ``coincidence'' because we have no intuitive explanation for this correspondence.}
the transmission probabilities of the evanescent modes are the same as those of diffusive modes in a disordered piece of metal with the same conductance \cite{Two06}. We will return to this ``pseudo-diffusive'' dynamics in Sec.\ \ref{Josephson}, when we describe how supercurrent flows through undoped ballistic graphene in the same way as it does through a disordered metal.

\begin{figure}[tb]
\includegraphics[width=0.9\linewidth]{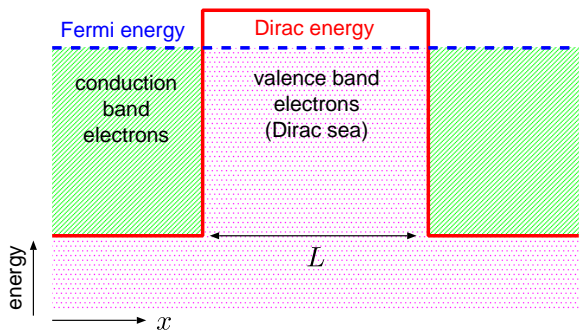}
\caption{\label{potential_b}
Electrostatic potential profile (red solid line) producing two heavily doped graphene regions at the left and right and a weakly doped region (length $L$) at the center. The central region is undoped when the Fermi energy (blue dashed line) coincides with the energy of the Dirac point. Electrical conduction then proceeds via evanescent (= exponentially decaying) modes.
}
\end{figure}

In preparation of that discussion, we examine here in a bit more detail the transmission of evanescent modes through undoped graphene \cite{Kat06b,Two06}. Because the wave length at the Dirac point is infinitely long, the detailed shape of the electrostatic potential profile at the interface between the metal contacts and the graphene sheet is not very important. We model it by the rectangular potential shown in Fig.\ \ref{potential_b}. The contact area is modeled by heavily doped graphene [for more microscopic models, see \textcite{Sch07,Rob07,Bla07}]. The Fermi level in Fig.\ \ref{potential_b} lies in the conduction band in the contact areas at the left and right and in the valence band in the central region. Conduction in this situation occurs via interband (Klein) tunneling, from conduction band to valence band, and we will have much more to say about that in Sec.\ \ref{Klein}.

The special situation we are interested in here is when the Fermi energy coincides with the energy of the Dirac point in the central region. At that energy interband tunneling goes over into intraband tunneling. For $W/L\gg 1$ we do not need to know the individual transmission probabilities of the evanescent modes (which will depend on the boundary condition at $y=0,W$), but it suffices to know how many modes $\rho(T)dT$ (counting all degeneracy factors) there are with transmission probabilities in the interval $(T,T+dT)$. The result is
\begin{equation}
\rho(T)=\frac{g}{2T\sqrt{1-T}},\;\;g=\frac{4W}{\pi L},\label{NTresult}
\end{equation}
with $g$ the conductance in units of $e^{2}/h$. We call the dynamics pseudo-diffusive because the distribution \eqref{NTresult} happens to coincide with the known distribution \cite{Dor84} for diffusion modes\footnote{
The $T$'s for diffusion modes are the eigenvalues of the transmission matrix product $tt^{\dagger}$. The distribution $\rho(T)$ for diffusion modes has a cutoff at exponentially small $T\simeq\exp(-2L/l)$, with $l$ the mean free path \cite{Bee97}. The distribution \eqref{NTresult} for evanescent modes has a cutoff at $\exp(-4\pi L/\lambda'_{F})$, with $\lambda'_{F}$ the Fermi wave length in the heavily doped regions. In either case the cutoff is irrelevant for transport properties.}
in a disordered metal having the same dimensionless conductance $g\gg 1$.

The value for $g$ in Eq.\ \eqref{NTresult} has been confirmed experimentally by \textcite{Mia07}. To test for the bimodal shape of the distribution $\rho(T)$ one would need to measure the shot noise at the Dirac point. The Fano factor (ratio of shot noise power and mean current) should equal \cite{Two06}
\begin{equation}
F=1-\frac{\int_{0}^{1}T^{2}\rho(T)\,dT}{\int_{0}^{1}T\rho(T)\,dT}=\frac{1}{3},\label{Fresult}
\end{equation}
just as in a disordered metal \cite{Bee92b}. This $1/3$-Fano factor has now been confirmed experimentally as well \cite{Dan07}.

\section{\label{Andreev} Andreev reflection}

\subsection{\label{ehconversion} Electron-hole conversion}

\begin{figure}[tb]
\includegraphics[width=0.9\linewidth]{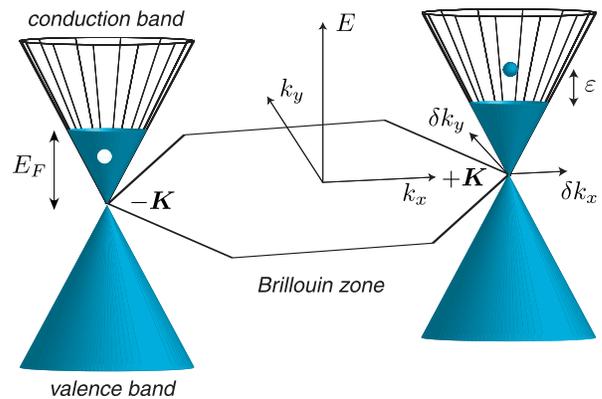}
\caption{\label{conical_point}
Electron and hole excitations in the conical band structure of graphene (filled and empty circles at energies $E_{F}\pm\varepsilon$), converted into each other by Andreev reflection at a superconductor.
}
\end{figure}

Andreev reflection is the conversion of electron into hole excitations by the superconducting pair potential \cite{And64}. The process is illustrated in Fig.\ \ref{conical_point} for the band structure of graphene. The electron excitation is a filled state at energy $\varepsilon$ above the Fermi energy $E_{F}$, while the hole excitation is an empty state at $\varepsilon$ below $E_{F}$. The excitation energy $\varepsilon$ is the same, so that Andreev reflection is an elastic process. Since electron and hole have opposite charge $\pm e$, a charge of $2e$ is lost in the conversion process. This missing charge is absorbed by the superconductor as a Cooper pair. For $\varepsilon$ below the superconducting gap $\Delta$ electrons can enter only pairwise into the superconductor, and the Andreev reflected hole is the empty state left behind by the electron that is paired with the incident electron to form a Cooper pair.

The electron and hole in Fig.\ \ref{conical_point} are taken from opposite corners $\pm{\bm K}$ of the Brillouin zone, in order to allow the Cooper pair to carry zero total momentum. This corresponds to the case of $s$-wave pairing, common in conventional (low-temperature) superconductors. Andreev reflection in graphene therefore switches the valleys \cite{Bee06}. The switching of valleys by Andreev reflection due to $s$-wave pairing in the superconductor is analogous to the switching of spin bands due to singlet pairing. The latter can be detected by producing a spin polarization in the normal metal \cite{Jon95}. Analogously, the former can be detected by producing a valley polarization in graphene, as we will discuss in Sec.\ \ref{QHE}.

The electron and hole in Fig.\ \ref{conical_point} are both from the conduction band. This {\em intraband\/} Andreev reflection applies if $\varepsilon<E_{F}$. For $\varepsilon>E_{F}$ the hole is an empty state in the valence band, rather than in the conduction band. In undoped graphene, when $E_{F}=0$, Andreev reflection is {\em interband\/} at all excitation energies. Interband Andreev reflection does not exist in usual metals, having an excitation gap $\gg\Delta$ between conduction and valence band. The peculiar differences between intraband and interband Andreev reflection are explained in the next subsection.

\subsection{\label{retro}Retro-reflection vs. specular reflection}

\begin{figure}[tb]
\includegraphics[width=0.8\linewidth]{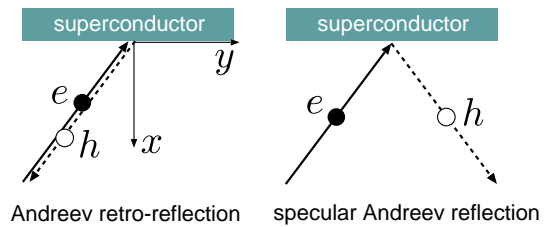}
\caption{\label{NSreflections}
The left panel shows Andreev retro-reflection at the interface between a normal metal and a superconductor. Arrows indicate the direction of the velocity and solid or dashed lines distinguish whether the particle is a negatively charged electron ($e$) or a positively charged hole ($h$). The right panel shows the specular Andreev reflection at the interface between undoped graphene and a superconductor. \cite{Bee06}
}
\end{figure}

\textcite{And64} discovered that the electron-hole conversion at a superconductor is associated with {\em retro-reflection\/} rather than {\em specular reflection}. Retro-reflection means that the reflected hole retraces the path of the incident electron (see Fig.\ \ref{NSreflections}, left panel) --- so all components of the velocity change sign. In undoped graphene, in contrast, Andreev reflection is specular (right panel) --- so only the component perpendicular to the interface changes sign \cite{Bee06}. 

\begin{figure}[tb]
\includegraphics[width=1\linewidth]{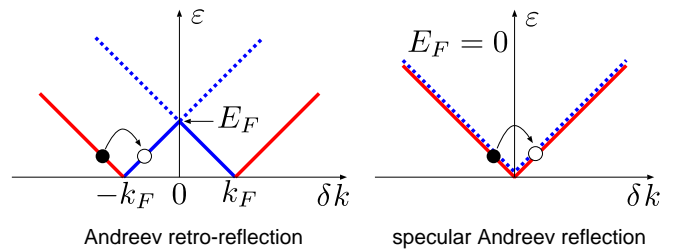}
\caption{\label{NSdispersions}
Dispersion relation \eqref{epsvk} in graphene for two values of the Fermi energy $E_{F}=\hbar vk_{F}$, for the case of normal incidence ($\delta k_{y}=0$, $\delta k_{x}\equiv \delta k$). Red lines indicate electron excitations (filled states above the Fermi level, from one valley), while blue lines indicate hole excitations (empty states below the Fermi level, from the other valley). Solid and dotted lines distinguish the conduction and valence bands, respectively. The electron-hole conversion upon reflection at a superconductor is indicated by the arrows. Specular Andreev reflection (right panel) happens if an electron in the conduction band is converted into a hole in the valence band. In the usual case (left panel), electron and hole both lie in the conduction band. \cite{Bee06}
}
\end{figure}

Inspection of the dispersion relation shows why intraband Andreev reflection leads to retro-reflection, while interband Andreev reflection leads to specular reflection. The linear dispersion relation \eqref{Evk} in graphene may be rewritten in terms of the excitation energy $\varepsilon=|E-E_{F}|$,
\begin{equation}
\varepsilon=|E_{F}\pm\hbar v(\delta k_{x}^{2}+\delta k_{y}^{2})^{1/2}|. \label{epsvk}
\end{equation}
The $\pm$ sign distinguishes excitations in the conduction and in the valence band. Let the interface with the superconductor be at $x=0$ and the electron approach the interface from $x>0$. Since $\delta k_{y}$ and $\varepsilon$ are conserved upon reflection, the reflected state is a superposition of the four $\delta k_{x}$-values that solve Eq.\ \eqref{epsvk} at given $\delta k_{y}$ and $\varepsilon$. The derivative $\hbar^{-1}d\varepsilon/d\delta k_{x}$ is the expectation value $v_{x}$ of the velocity in the $x$-direction, so the reflected state contains only the two $\delta k_{x}$-values having a positive slope. One of these two allowed $\delta k_{x}$-values is an electron excitation, the other a hole excitation. As illustrated in Fig.\ \ref{NSdispersions}, the reflected hole may be either an empty state in the conduction band (for $\varepsilon<E_{F}$) or an empty state in the valence band ($\varepsilon>E_{F}$). A conduction-band hole moves opposite to its wave vector, so $v_{y}$ changes sign as well as $v_{x}$ (retro-reflection). A valence-band hole, in contrast, moves in the same direction as its wave vector, so $v_{y}$ remains unchanged and only $v_{x}$ changes sign (specular reflection).

\begin{figure}[tb]
\includegraphics[width=0.8\linewidth]{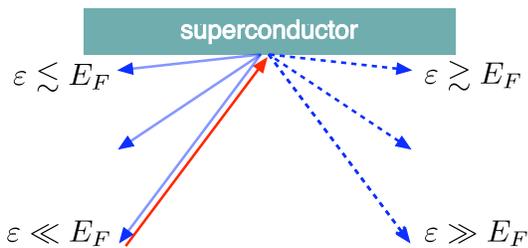}
\caption{\label{reflection_angles}
Trajectories of an incident electron (red) and the Andreev reflected hole (blue), for different excitation energies $\varepsilon$ relative to the Fermi energy $E_{F}$, at fixed angle of incidence. For $\varepsilon<E_{F}$ the hole is in the conduction band (solid lines), while for $\varepsilon>E_{F}$ the hole is in the valence band (dashed lines). The reflected trajectories rotate clockwise with increasing $\varepsilon$,  jumping by $180^{\circ}$ when $\varepsilon=E_{F}$.
}
\end{figure}

The transition from retro-reflection to specular reflection as $\varepsilon$ increases beyond $E_{F}$ is illustrated in Fig.\ \ref{reflection_angles}. The reflection angle $\alpha_{\rm out}$ (measured relative to the normal) first becomes greater than the angle of incidence $\alpha_{\rm in}$, then jumps from $+90^{\circ}$ to $-90^{\circ}$ at $\varepsilon=E_{F}$, and finally approaches $-\alpha_{\rm in}$ when $\varepsilon\gg E_{F}$.

\begin{figure}[tb]
\includegraphics[width=0.9\linewidth]{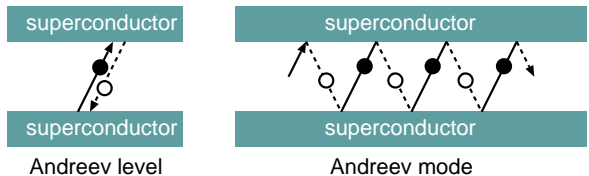}
\caption{\label{Andreev_mode}
The transition from retro-reflection to specular Andreev reflection in a graphene channel with superconducting boundaries induces a transition from a localized level (left) to a propagating mode (right). The latter state contributes to thermal transport along the channel, but not to electrical transport. \cite{Tit07}
}
\end{figure}

As shown in Fig.\ \ref{Andreev_mode}, specular Andreev reflection creates charge-neutral propagating modes along an undoped graphene channel with superconducting boundaries \cite{Tit07}. In contrast, retro-reflection creates bound states known as Andreev levels \cite{And64,Kul70}. The propagating ``Andreev modes'' contribute to the thermal conductance along the graphene channel in a way which is sensitive to the superconducting phase difference across the channel. They may also be used to carry a charge-neutral spin current along the channel \cite{Gre07}. We will return to this geometry in Sec.\ \ref{Josephson}, when we consider the current across the channel (from one superconductor to the other) --- rather than along the channel.

\subsection{\label{DBdG} Dirac-Bogoliubov-de Gennes equation}

So far our discussion of Andreev reflection in graphene has been semiclassical, in terms of electron and hole trajectories. Quantum mechanically, the coupling of electron and hole wave functions $\Psi_{e}$ and $\Psi_{h}$ is described by the Dirac-Bogoliubov-de Gennes (DBdG) equation \cite{Bee06},
\begin{equation}
\begin{pmatrix}
H(\bm{A})-E_{F}&\Delta(\sigma_{0}\otimes\tau_{0})\\
\Delta^{\ast}(\sigma_{0}\otimes\tau_{0})&E_{F}-H(-\bm{A})
\end{pmatrix}
\begin{pmatrix}
\Psi_{e}\\ \Psi_{h} 
\end{pmatrix}=
\varepsilon
\begin{pmatrix}
\Psi_{e}\\ \Psi_{h}
\end{pmatrix}.\label{BdGeq}
\end{equation}
The complex pair potential $\Delta=\Delta_{0}e^{i\Phi}$ is nonzero only in the superconducting region, where it couples the time-reversed states
\begin{equation}
\begin{split}
\Psi_{e}&=(\Psi_{A},\Psi_{B},-\Psi'_{B},\Psi'_{A}),\\
\Psi_{h}&={\cal T}\Psi_{e}=({\Psi'}^{\ast}_{A},{\Psi'}^{\ast}_{B},-\Psi^{\ast}_{B},\Psi^{\ast}_{A}).
\end{split}
\label{Psiehdef}
\end{equation}
The boundary condition for the DBdG equation at the edges of the graphene sheet is given by the same Eq.\ \eqref{boundarycondition} for both $\Psi_{e}$ and $\Psi_{h}$,
\begin{equation}
\Psi_{e}=M\Psi_{e},\;\;\Psi_{h}=M\Psi_{h},\label{boundaryconditioneh}
\end{equation}
since we are assuming that $M$ commutes with ${\cal T}$.

In the normal region $\Delta\equiv 0$, so that there $\Psi_{e}$ and $\Psi_{h}$ satisfy the uncoupled equations
\begin{equation}
\begin{split}
H(\bm{A})\Psi_{e}&=(E_{F}+\varepsilon)\Psi_{e},\\
H(-\bm{A})\Psi_{h}&=(E_{F}-\varepsilon)\Psi_{h}.
\end{split}
\label{PsiehN}
\end{equation}
Andreev reflection at the normal-superconductor (NS) interface couples $\Psi_{e}$ to $\Psi_{h}$. This coupling may be described by means of a boundary condition at the NS interface for the wave function in the normal region \cite{Tit06},
\begin{equation}
\Psi_{h}=e^{-i\Phi}e^{-i\beta\,\bm{n}_{B}\cdot\bm{\sigma}}\otimes\tau_{0}\Psi_{e},\label{Psiehrelation}
\end{equation}
where $\beta=\arccos(\varepsilon/\Delta_{0})\in(0,\pi/2)$ (assuming $\varepsilon<\Delta_{0}$). The unit vector $\bm{n}_{B}$ is perpendicular to the NS interface, pointing from N to S. By substituting the definition \eqref{Psiehdef} of $\Psi_{e}$ and $\Psi_{h}$ we see that the boundary condition \eqref{Psiehrelation} couples electron excitations in one valley to hole excitations in the other valley (in accord with the description of Andreev reflection given in Sec.\ \ref{ehconversion}). In contrast, the boundary condition \eqref{boundaryconditioneh} at the edges of the graphene sheet does not couple $\Psi_{e}$ and $\Psi_{h}$.

The relation (\ref{Psiehrelation}) follows from the DBdG equation \eqref{BdGeq} under three assumptions characterizing an ``ideal'' NS interface:
\begin{itemize}
\item
The Fermi wave length $\lambda'_{F}$ in S is sufficiently small that $\lambda'_{F}\ll\xi,\lambda_{F}$, where $\lambda_{F}=hv/E_{F}$ is the Fermi wave length in N and $\xi=\hbar v/\Delta_{0}$ is the superconducting coherence length.
\item
The interface is smooth and impurity free on the scale of $\xi$.
\item
There is no lattice mismatch at the NS interface, so the honeycomb lattice of graphene is unperturbed at the boundary. 
\end{itemize}
The absence of lattice mismatch might be satisfied by depositing the superconductor on top of a heavily doped region of graphene. As in the case of a semiconductor two-dimensional electron gas \cite{Vol95,Fag05}, we expect that such an extended superconducting contact can be effectively described by a pair potential $\Delta$ in the $x$-$y$ plane (even though graphene by itself is not superconducting).

At normal incidence $\Psi_{e}$ and $\Psi_{h}$ are eigenstates of $\bm{n}_{B}\cdot\bm{\sigma}$, so the boundary condition \eqref{Psiehrelation} implies that $\Psi_{h}=\Psi_{e}\times{}$ a phase factor and the electron-hole conversion happens with unit probability. This is entirely different from usual NS junctions, where Andreev reflection is suppressed at any angle of incidence if the Fermi wave lengths at the two sides of the interface are very different.

\subsection{\label{Josephson} Josephson junctions}

The boundary condition \eqref{Psiehrelation} at a normal-superconducting interface depends on the phase $\Phi$ of the superconductor, although this dependence is unobservable if there is only a single superconductor. A Josephson junction is a junction between two superconductors with a controllable phase difference $\phi=\Phi_{1}-\Phi_{2}$. A current $I(\phi)$ flows from one superconductor to the other if $\phi\neq 0$. The current flows in equilibrium, so it is a dissipationless supercurrent. This is the Josephson effect \cite{Jos64}. Since $I$ is $2\pi$-periodic in $\phi$, there exists a maximal supercurrent $I_{c}$ that can flow between the superconductors. This is called the critical current of the Josephson junction.

There is a thermodynamic relation \cite{And63}
\begin{equation}
I=\frac{2e}{\hbar}\frac{dF}{d\phi}\label{Iphi1}
\end{equation}
between the supercurrent $I$ and the derivative of the free energy $F$ with respect to the superconducting phase difference. The free energy can in turn be related to the excitation spectrum, which itself follows from the DBdG equation. At zero temperature and in the short-junction limit (separation $L$ of the two NS interfaces $\ll\xi$) the resulting relation is \cite{Bee92}
\begin{equation}
I=-\frac{2e}{\hbar}\sum_{n}\frac{d}{d\phi}\varepsilon_{n}(\phi),\label{Iphi2}
\end{equation}
with $\varepsilon_{n}<\Delta_{0}$ the energy of a (spin-degenerate) bound state in the Josephson junction.

\begin{figure}[tb]
\includegraphics[width=0.9\linewidth]{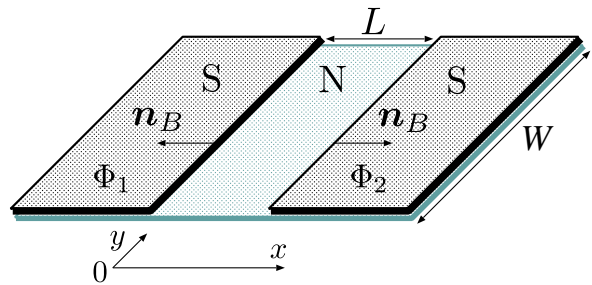}
\caption{\label{Jojunction}
Josephson junction, formed by a graphene layer (N) with two superconducting electrodes (S) a distance $L$ apart, having a phase difference $\phi=\Phi_{1}-\Phi_{2}$. \cite{Tit06}
}
\end{figure}

To calculate the supercurrent (in zero magnetic field) one therefore needs to solve the two eigenvalue equations \eqref{PsiehN} (with $\bm{A}=0$) in the strip $0<x<L$, $0<y<W$ (see Fig.\ \ref{Jojunction}). At $x=0,L$ there is the phase-dependent boundary condition \eqref{Psiehrelation}, which couples $\Psi_{e}$ to $\Psi_{h}$, while the boundary condition \eqref{boundarycondition} at $y=0,W$ is phase-independent and does not couple $\Psi_{e}$ to $\Psi_{h}$. 

\begin{figure}[tb]
\includegraphics[width=0.8\linewidth]{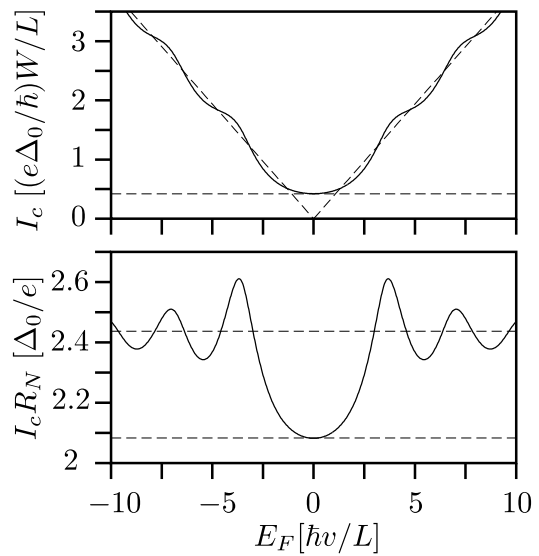}
\caption{\label{IcRn}
Critical current $I_{c}$ and $I_{c}R_{N}$ product of a ballistic Josephson junction (length $L$ short compared to the width $W$ and superconducting coherence length $\xi$), as a function of the Fermi energy $E_{F}$ in the normal region. Small and large $|E_{F}|$ asymptotes are indicated by dashed lines. \cite{Tit06}
}
\end{figure}

The result of this calculation \cite{Tit06} is that the critical current is given, up to numerical coefficients of order unity, by
\begin{equation}
I_{c}\simeq\frac{e\Delta_{0}}{\hbar}\,\max(W/L,W/\lambda_{F}).\label{Icresult}
\end{equation}
(The dependence on the boundary condition at $y=0,W$ can be neglected under the assumption $L\ll W$ of a short and wide junction.) At the Dirac point $E_{F}=0$ one has $\lambda_{F}\rightarrow\infty$, so the critical current reaches its minimal value $\simeq(e\Delta_{0}/\hbar)\times W/L$ (see Fig.\ \ref{IcRn}, upper panel). Instead of being independent of the length $L$ of the junction, as expected for a short ballistic Josephson junction, the critical current at the Dirac point has the diffusion-like scaling $\propto 1/L$. This is another manifestation of the ``pseudo-diffusive'' dynamics discussed in Sec.\ \ref{pseudodif}.

\begin{figure}[tb]
\includegraphics[width=.7\linewidth]{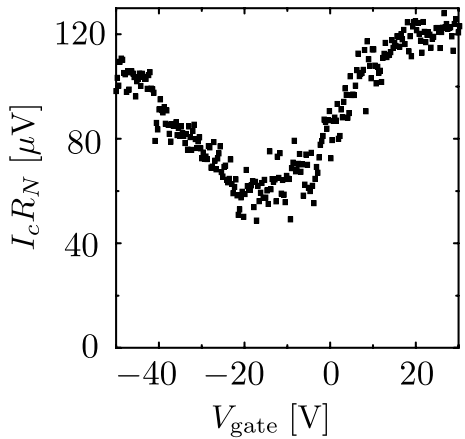}
\caption{\label{IcRn_exp}
Product of the critical current $I_{c}$ and the normal state resistance $R_{N}$ versus gate voltage $V_{\rm gate}$, measured at $T=30\,{\rm mK}$ in the Josephson junction of Fig.\ \ref{SNSmonolayer}. The carrier density in the graphene layer is linearly proportional to $V_{\rm gate}$, while the Fermi energy $E_{F}\propto\sqrt V_{\rm gate}$. The resistance $R_{N}$ is measured in the presence of a small magnetic field to drive the electrodes in the normal state. \cite{Hee07}
}
\end{figure}

Since the normal-state resistance scales as \cite{Kat06b,Two06}
\begin{equation}
1/R_{N}\equiv G_{N}\simeq\frac{e^{2}}{h}\,\max(W/L,W/\lambda_{F}),\label{RNresult}
\end{equation}
the theory predicts that the $I_{c}R_{N}$ product remains of order $\Delta_{0}/e$ (up to a numerical prefactor) as the Fermi level passes through the Dirac point (Fig.\ \ref{IcRn}, lower panel). The experimental result of \textcite{Hee07} for the Josephson junction of Fig.\ \ref{SNSmonolayer}, shown in Fig.\ \ref{IcRn_exp}, is qualitatively similar to the theoretical prediction, but there are significant quantitative differences: The experimental $I_{c}R_{N}$ product at the Dirac point is about $60\,\mu{\rm V}\approx 0.5\,\Delta_{0}/e$, more than twice the theoretical prediction, and the increase at higher carrier densities is much larger than predicted. It is quite likely that disorder in the experimental sample, which is not included in the calculation, is responsible for these differences \cite{Du07}.

\subsection{\label{furtherNS} Further reading}

In the spirit of a Colloquium, we have only discussed the basic physics of Andreev reflection in graphene. In this subsection we give some pointers to the literature on other aspects of this topic.

The pseudo-diffusive dynamics at the Dirac point, discussed in Sec.\ \ref{Josephson} in connection with the critical current $I_{c}$ of an SNS junction, extends to the entire current-phase relationship $I(\phi)$ in equilibrium \cite{Tit06}, as well as to the dissipative current out of equilibrium \cite{Cue06}. In each case, a short and wide strip of undoped ballistic graphene (length $L$ short compared to width $W$ and superconducting coherence length $\xi$)\footnote{
This short-junction limit is essential: Pseudo-diffusive dynamics in SNS junctions breaks down if $L$ becomes larger than $W$ \cite{Mog06,Gon07} or if $L$ becomes larger than $\xi$ \cite{Tit07}. A tunnel barrier \cite{Mai07} or \textit{p-n} junction \cite{Oss07} in the normal region also spoils the pseudo-diffusive analogy.
}
behaves as a disordered metal having the same normal-state conductance $G_{N}$.

Pseudo-diffusive dynamics also governs the conductance $G_{NS}$ through a ballistic graphene strip ($L\ll W$) having a single superconducting contact \cite{Akh07ba,Pra07}, in the sense that the ratio $G_{NS}/G_{N}$ at the Dirac point is the same as for a disordered metal. The correspondence holds only for voltages $V$ small compared to $\hbar v/L$. At larger voltages the current-voltage characteristic of a ballistic NS junction in graphene has unusual features \cite{Bee06,Bha06,Bha07} --- without a diffusive analogue. These have been studied experimentally by \textcite{Sha07} and \textcite{Mia07}. Similarly unusual $I$-$V$ characteristics have been predicted in bilayer graphene \cite{Lud07}.

The Dirac-Bogoliubov-de Gennes equation of Sec.\ \ref{DBdG} assumes isotropic ($s$-wave) pairing in the superconductor. The equation may be readily modified to the case of anisotropic ($d$-wave) pairing, relevant for NS contacts between graphene and a high-temperature superconductor. The conductance in the two cases has been compared by \textcite{Lin07}. Another modification, studied by \textcite{Weh07}, is to include electrical or magnetic scattering potentials in the superconducting region.

More exotic ($p_{x}+ip_{y}$ or $d_{x^{2}-y^{2}}+id_{xy}$) pairings may be possible \cite{Uch07,Jia07} if graphene could be chemically modified to become an intrinsic superconductor (rather than having the superconductivity induced by the proximity effect). \textcite{Gha07} have argued that the special topological properties (nonabelian statistics) of vortices in a superconductor with $p_{x}+ip_{y}$ pairing apply as well to the $s$-wave DBdG equation \eqref{BdGeq}, if the superconductivity can be induced in undoped graphene.

The idealized model of the NS interface discussed in Sec.\ \ref{DBdG} can be much improved, in particular to include the effects of lattice mismatch and a selfconsistent calculation of the induced pair potential. Some numerical \cite{Wak03} and analytical \cite{Tka07} work goes in this direction.

\section{\label{Klein} Klein tunneling}

\subsection{\label{backscattering} Absence of backscattering}

The massless carriers in graphene respond quite differently to an electric field than ordinary massive carriers. Because the magnitude $v$ of the carrier velocity is independent of the energy, an electron moving along the field lines cannot be backscattered --- since that would require $v=0$ at the turning point. The absence of backscattering was discovered by \textcite{And98} in carbon nanotubes, where it is responsible for the high conductivity in the presence of disorder. The two-dimensional nature of the dynamics in graphene introduces some new aspects.

\begin{figure}[tb]
\includegraphics[width=0.8\linewidth]{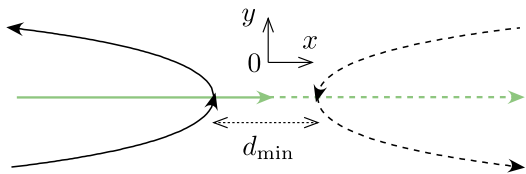}
\caption{\label{Dirac_trajectories}
Classical trajectories of an electron in the presence of a uniform electric field in the $x$-direction. All three trajectories are at the same energy, only the component $p_{y}$ of the momentum transverse to the field lines is varied. The two black trajectories are for $p_{y}>0$, while the green trajectory is for $p_{y}=0$. The electron is in the conduction band of graphene for $x<0$ (solid trajectories, velocity parallel to momentum) and in the valence band for $x>0$ (dashed trajectories, velocity antiparallel to momentum). Solid and dashed trajectories are coupled by Klein tunneling.
}
\end{figure}

\begin{figure}[tb]
\includegraphics[width=0.9\linewidth]{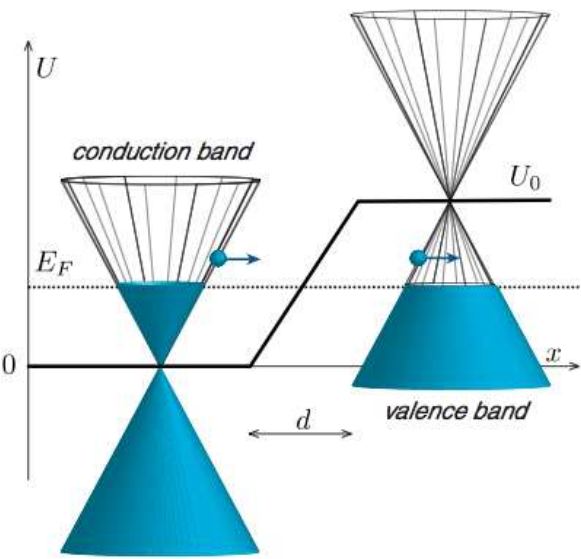}
\caption{\label{Klein_tunneling}
Band structure of a single valley at two sides of a potential step (height $U_{0}$, width $d$). The equilibrium Fermi energy $E_{F}$ is the same at both sides, so that for $U_{0}>E_{F}$ an electron just above the Fermi level is in the conduction band at one side and in the valence band at the other side. Blue arrows indicate the electron velocity, which is parallel to the wave vector (or momentum) in the conduction band (left) and antiparallel in the valence band (right).
}
\end{figure}

Electron trajectories in the linear electrostatic potential $U(x)={\cal F}x$ are shown in Fig.\ \ref{Dirac_trajectories}. The trajectories are deflected by the electric field for $p_{y}\neq 0$, but for $p_{y}=0$ no backscattering occurs. The electron is able to propagate through an infinitely high potential barrier because it makes a transition from the conduction band to the valence band (see Fig.\ \ref{Klein_tunneling}). In this transition its dynamics changes from electron-like to hole-like in the following sense:

The equation of motion
\begin{equation}
\frac{d\bm{r}}{dt}\equiv\frac{\partial E}{\partial\bm{p}}=\frac{v^{2}\bm{p}}{E-U},
\end{equation} 
at energy $E$ with $v^{2}|\bm{p}|^{2}=(E-U)^{2}$, implies that the velocity $\bm{v}=d\bm{r}/dt$ of the electron is parallel to the momentum when it is in the conduction band ($U<E$) and antiparallel when it is in the valence band ($U>E$). States with $\bm{v}$ parallel to $\bm{p}$ are called electron-like and states with $\bm{v}$ antiparallel to $\bm{p}$ are called hole-like. By making the transition from electron-like to hole-like dynamics, the electron can continue to move in the same direction even as its momentum along the field lines goes through zero and changes sign.

In classical mechanics, backscattering is only avoided for $p_{y}=0$ (so only if the electron moves along the field lines). In quantum mechanics an electron can tunnel from the conduction into the valence band, thereby avoiding backscattering, also for a small but nonzero $p_{y}$. Such tunneling from an electron-like to a hole-like state is called interband tunneling \cite{Aro67,Wei67,Kan69} or Klein tunneling \cite{Kat06}, because of an analogous effect in relativistic quantum mechanics \cite{Kle29}.\footnote{
Klein tunneling is considered paradoxical in the relativistic context \cite{Cal99}, because the hole-like states into which the electron tunnels are unphysical antiparticle states in the Dirac sea. There is no paradox in the context of graphene, where the role of the Dirac sea is played by the valence band (see Fig.\ \ref{potential_b}).} 

The probability of Klein tunneling of a relativistic electron in a uniform electric field was calculated by \textcite{Sau31} --- with an exponentially small result due to the finite electron mass. The case of massless particles, relevant for graphene, was considered by \textcite{Che06}. Pairs of electron-like and hole-like trajectories at the same $E$ and $p_{y}$ have turning points at a distance $d_{\rm min}=2v|p_{y}|/{\cal F}$. The tunneling probability has an exponential dependence on this separation,\footnote{
The asymptotic result \eqref{TU0result}, derived by \textcite{Che06} and \textcite{And07}, should follow from the general Kummer-function formula of \textcite{Sau31} upon substitution of the electron mass $m$ by $p_{y}/v$ and taking the limit \eqref{pinoutcondition}. The asymptotic limit taken by Sauter corresponds to the opposite regime $|p_{y}|\gg |p^{\rm in}_{x}|,|p^{\rm out}_{x}|,\sqrt{\hbar {\cal F}/v}$ in which $T(p_{y})$ is exponentially small.}
\begin{equation}
T(p_{y})=\exp(-\pi |p_{y}|d_{\rm min}/2\hbar)=\exp(-\pi v p_{y}^{2}/\hbar {\cal F}),\label{TU0result}
\end{equation}
provided that the longitudinal momentum $p^{\rm in}_{x}$ at $x\rightarrow-\infty$ and $p^{\rm out}_{x}$ at $x\rightarrow\infty$ (where the electric field is assumed to vanish) is sufficiently large:
\begin{equation}
|p^{\rm in}_{x}|,|p^{\rm out}_{x}|\gg|p_{y}|,\sqrt{\hbar {\cal F}/v}.\label{pinoutcondition}
\end{equation}

Transmission resonances occur when a \textit{p-n} interface is in series with an \textit{n-p} interface, forming a \textit{p-n-p} or \textit{n-p-n} junction \cite{Kat06,Mil06,Sil07,Mil07b}. The unit transmission at $p_{y}=0$ forbids transmission resonances at normal incidence, in marked contrast with conventional resonant tunneling through a double-barrier junction.

\subsection{\label{pn} Bipolar junctions}

\begin{figure}[tb]
\includegraphics[width=0.8\linewidth]{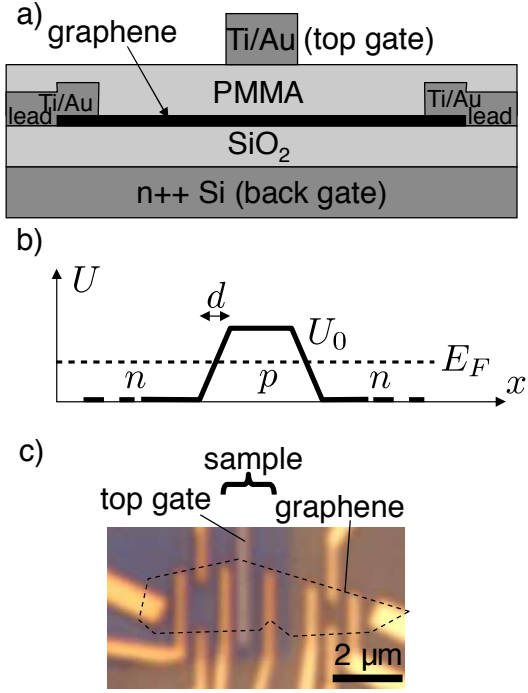}
\caption{\label{Huard}
\textit{n-p-n} junction in graphene: a) cross-sectional view of the device. b) electrostatic potential profile $U(x)$
along the cross-section of a). The combination of a positive voltage on the back gate and a negative voltage on the top gate produces a central \textit{p}-doped region flanked by two \textit{n}-doped regions. c) Optical image of the device. The barely visible graphene flake is outlined with a dashed line and the dielectric layer of PMMA appears as a blue shadow. \cite{Hua07}
}
\end{figure}

Klein tunneling is the mechanism for electrical conduction through the interface between \textit{p}-doped and \textit{n}-doped graphene. Such a bipolar junction is illustrated in Fig.\ \ref{Huard} \cite{Hua07}. A top gate creates an electrostatic potential barrier, so that the Fermi level lies in the valence band inside the barrier (\textit{p}-doped region) and in the conduction band outside the barrier (\textit{n}-doped region). The carrier density $n_{\rm carrier}$ is the same in the \textit{n} and \textit{p} regions when the Fermi energy $E_{F}$ is half the barrier height $U_{0}$. In this case the Fermi momenta $p_{F}\equiv\hbar k_{F}$ in both \textit{n} and \textit{p} regions are given by $p_{F}= U_{0}/2v={\cal F}d/2v$, with $d$ the width of the \textit{n-p} and $\textit{p-n}$ interfaces and ${\cal F}=U_{0}/d$ the electric field (up to a factor of electron charge) in that interface region.\footnote{
This assumption of a constant electric field in the interface region requires perfect screening by the carriers in graphene of the electric field produced by the gate. The lack of screening at the \textit{p-n} interface due to the vanishing carrier density enhances the local electric field by a factor $(e^{2}k_{F}d/\kappa\hbar v)^{1/3}$, with $\kappa$ the dielectric constant \cite{Zha07}. The value of $\kappa$ can be as low as 2.4 for a ${\rm SiO}_{2}$ substrate and as large as 80 for graphene on water.}

The width $d$ is of the order of the separation between the graphene layer and the top gate. \textcite{Hua07} estimate $d\approx 80\,{\rm nm}$ for their device. The Fermi wave vector $k_{F}=\sqrt{\pi n_{\rm carrier}}$ is $\gtrsim 10^{-1}\,{\rm nm}^{-1}$, for typical carrier densities of $n_{\rm carrier}\gtrsim 10^{12}\,{\rm cm}^{-2}$. Since under these conditions $k_{F}d> 1$ the \textit{p-n} and \textit{n-p} interfaces are smooth on the scale of the Fermi wave length. This is the regime of applicability of the expression \eqref{TU0result} for the Klein tunneling probability, since the condition \eqref{pinoutcondition} of large longitudinal momentum can be rewritten as
\begin{equation}
p_{F}\gg|p_{y}|,\hbar/d.\label{pFcondition}
\end{equation}

The conductance $G_{\textit{p-n}}$ of a \textit{p-n} interface follows by integration of Eq.\ \eqref{TU0result} over the transverse momenta, with the result \cite{Che06}
\begin{equation}
G_{\textit{p-n}}=\frac{4e^{2}}{h}\frac{W}{2\pi\hbar}\int_{-\infty}^{\infty}dp_{y}\,T(p_{y})=\frac{4e^{2}}{h}\frac{W}{2\pi}\sqrt{\frac{{\cal F}}{\hbar v}}.
\label{Gpnresult}
\end{equation}
The factor of $4$ accounts for the twofold spin and valley degeneracy and $W$ is the transverse dimension of the interface. The integration range may be extended to $\pm\infty$ because $T(p_{y})$ is vanishingly small for $|p_{y}|$ larger than $\sqrt{\hbar{\cal F}/v}\simeq p_{F}/\sqrt{k_{F}d}\ll p_{F}$. Notice that $G_{\textit{p-n}}$ is smaller than the ballistic conductance $G_{\rm ballistic}=(4e^{2}/h)k_{F}W/\pi$ by the same factor $\sqrt{k_{F}d}$ that characterizes the smoothness of the interface.

\subsection{\label{Beffects} Magnetic field effects}

\begin{figure}[tb]
\includegraphics[width=0.8\linewidth]{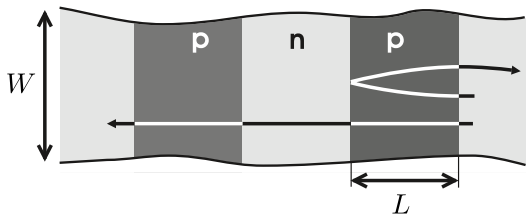}
\caption{\label{Cheianov_Falko}
Two trajectories in a \textit{p-n-p} junction, the lower one (transmitted) in zero magnetic field and the upper one (reflected) in a small but nonzero field. Because only trajectories with an angle of order $1/\sqrt{k_{F}d}\ll 1$ around normal incidence are transmitted through the \textit{p-n} and \textit{n-p} interfaces, a relatively weak magnetic field suppresses the series conductance of the interfaces by bending the trajectories away from normal incidence. \cite{Che06}
}
\end{figure}

\textcite{Che06} have predicted that a relatively weak magnetic field $B\simeq(\hbar/e)\sqrt{k_{F}/L^{2}d}$ will suppress the conductance of an \textit{n-p-n} or \textit{p-n-p} junction (of length $L$) below the series conductance of the individual interfaces, as a consequence of the strong angular dependence of the transmission probability \eqref{TU0result}. The mechanism is illustrated in Fig.\ \ref{Cheianov_Falko}. The effect is not observed in the device of \textcite{Hua07}, presumably because of disorder. (See \textcite{Fog07} for a calculation of the conditions required for ballistic transport, which are only marginally met in existing experiments.)

The conductance of a single \textit{p-n} interface becomes magnetic field dependent on the much larger field scale $B_{\ast}={\cal F}/ev\simeq (\hbar/e)k_{F}/d$ at which the cyclotron radius $l_{\rm cycl}=\hbar k_{F}/eB$ becomes comparable to the width $d$ of the interface. \textcite{Shy07} have calculated that the angle of incidence $\theta_{\rm max}$ which is maximally transmitted rotates away from normal incidence to a value $\theta_{\rm max}=\pm\arcsin(B/B_{\ast})$. The effect on the conductance \eqref{Gpnresult} of the \textit{p-n} interface is a suppression by a factor $[1-(B/B_{\ast})^{2}]^{3/4}$.

\begin{figure}[tb]
\includegraphics[width=0.9\linewidth]{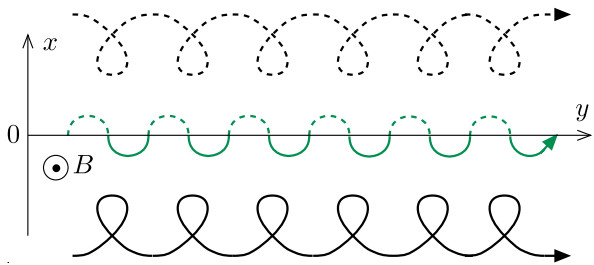}
\caption{\label{NSpn_B}
Electron trajectories along a \textit{p-n} interface in a magnetic field $B>B_{\ast}$ (when there is no transmission through the interface). The electron rotates in opposite directions in the conduction band (solid trajectories) and in the valence band (dashed). The trajectory centered at the interface (green) represents an ``ambipolar snake state''.
}
\end{figure}

For $B>B_{\ast}$ no transmission is possible through the \textit{p-n} interface. Instead, states in both the \textit{p} and \textit{n} regions propagate {\em parallel\/} to the interface \cite{Luk07,Mil07a}. The corresponding classical trajectories are illustrated in Fig.\ \ref{NSpn_B}. The direction of propagation along the interface is the same in both the \textit{p} and \textit{n} regions \cite{Aba07} --- while the direction of rotation is opposite. The snake-shaped trajectory centered at the interface has a mixed electron-hole character. This is the ambipolar analogue of the snake states that are known to exist in a nonhomogeneous magnetic field \cite{Mul92,Rak07,Gho07}. 

The conductance in the high-field regime $B>B_{\ast}$ is not fully suppressed, but it no longer scales with the width $W$ of the junction. We will calculate it in Sec.\ \ref{QHE}. 

\subsection{\label{furtherpn} Further reading}

Just as in Sec.\ \ref{furtherNS}, we mention here some papers for further reading on this topic.

Klein tunneling in a carbon bilayer differs fundamentally from Klein tunneling in a monolayer \cite{Kat06}. The bilayer still has a gapless spectrum [in the absence of a potential difference between the layers \cite{McC06a,Oht06}], so interband tunneling can happen with high probability. However, at normal incidence the probability is $0$ --- while it is $1$ in the monolayer. Although electrons in a bilayer are not massless, as they are in a monolayer, they still have a definite chirality (direction of motion tied to direction of pseudospin). Klein tunneling in a carbon bilayer is therefore different from interband tunneling in a gapless semiconductor. For example, the chirality forbids transmission resonances at normal incidence.

The perfect transmission at normal incidence in a monolayer is a robust effect with regard to the shape of the electrostatic potential profile at the \textit{p-n} interface (all that is needed is a potential which is smooth on the scale of the lattice constant). A time dependent electric field parallel to the interface, however, can suppress the transmission even at normal incidence \cite{Tra07,Fis07}. The suppression is strongest if the frequency $\omega$ of the radiation satisfies the resonance condition $\omega=2v|\bm{p}|/\hbar$ at some point in the interface region.

Bipolar junctions may appear naturally in disordered graphene, when the random electrostatic potential landscape produces alternating regions of \textit{p}-type and \textit{n}-type doping \cite{Mar07}. Classical percolation through such a random network of bipolar junctions has been studied by \textcite{Che07b}. At zero Fermi energy (when the areas of \textit{p}-type and \textit{n}-type doping are the same), the percolation length remains finite because of Klein tunneling.

Klein tunneling is also responsible for the finite life time of an electron state bound to a charged impurity in graphene \cite{Shy07b}. Such quasi-bound states exist for $\beta=Ze^{2}/\kappa\hbar v_{F}>1/2$, with $Ze$ the impurity charge. The discrete states exist in the conduction band near the impurity, but they are coupled by Klein tunneling to the continuum of states in the valence band away from the impurity. The resulting resonances (having width-to-energy ratio $e^{-2\pi \beta}$) may be observable by measuring the local density of states with a scanning probe. 

\section{\label{Analogies} Analogies}

In the previous two sections we have discussed NS and \textit{p-n} junctions separately. In this section we address the analogies between these two systems. Both involve the coupling of electron-like and hole-like states, either by the superconducting pair potential (in the NS junction) or by the electrostatic potential (in the \textit{p-n} junction). An obvious difference is that the two types of states lie {\em at the same side\/} of the NS interface but {\em at opposite sides\/} of the \textit{p-n} interface. The analogies, therefore, involve a reflection of the geometry along the interface \cite{Two07}.

\subsection{\label{mapping} Mapping between NS and \textit{p-n} junction
}

A precise mapping \cite{Bee07} between NS and \textit{p-n} junctions is possible under two conditions:
\begin{itemize}
\item The electrostatic potential $U$ in the \textit{p-n} junction is antisymmetric, $U(-x,y)=-U(x,y)$, with respect to the \textit{p-n} interface at $x=0$.
\item The NS interface may be described by the boundary condition \eqref{Psiehrelation} at $x=0$.
\end{itemize}
A uniform perpendicular magnetic field $B$ may or may not be present. Under these conditions a \textit{p-n} junction has the same excitation spectrum as an NS junction for $E_{F}=0$ and excitation energies $\varepsilon\ll\Delta_{0}$.

This correspondence follows from the fact that, if $\Psi$ is an eigenstate of the Dirac Hamiltonian \eqref{HDirac} of the \textit{p-n} junction with eigenvalue\footnote{
Since the spectrum of the \textit{p-n} junction is symmetric around zero energy, it suffices to consider energies $\varepsilon>0$.
}
$\varepsilon$, then we can construct an eigenstate $(\Psi_{e},\Psi_{h})$ in the normal part $x>0$ of the NS junction by 
\begin{equation}
\begin{split}
\Psi_{e}(x,y)&=\Psi(x,y),\\
\Psi_{h}(x,y)&=ie^{-i\Phi}(\sigma_{x}\otimes\tau_{0})\Psi(-x,y)\equiv {\cal P}\Psi(x,y).
\end{split}
\label{PsiehPsirelation}
\end{equation}
Here ${\cal P}=ie^{-i\Phi}(\sigma_{x}\otimes\tau_{0}){\cal R}$ with ${\cal R}$ the reflection operator ($x\mapsto -x$). Since $\Psi$ is continuous at $x=0$, the boundary condition \eqref{Psiehrelation} at the NS interface is automatically satisfied for $\varepsilon\ll\Delta_{0}$. Furthermore, from $H\Psi=\varepsilon\Psi$ and ${\cal P}H(\bm{A})=-H(-\bm{A}){\cal P}$ (with $\bm{A}=Bx\bm{\hat{y}}$) it follows that $\Psi_{e}$ and $\Psi_{h}$ satisfy the DBdG equation \eqref{PsiehN} in the normal region.

The applicability of the mapping may be extended in several ways: The \textit{p-n} junction may have boundaries described by the boundary condition $\Psi(\bm{r})=M(\bm{r})\Psi(\bm{r})$ for $\bm{r}$ at the boundary. (We may assume that this relation holds for all $\bm{r}$ by setting $M\equiv 1$ when $\bm{r}$ is not at the boundary.) The mapping to an NS junction still holds, provided that $M$ commutes with ${\cal P}$, which requires
\begin{equation}
(\sigma_{x}\otimes\tau_{0})M(x,y)=M(-x,y)(\sigma_{x}\otimes\tau_{0}). \label{MPrelation}
\end{equation}
This ensures that the transformed wave function \eqref{PsiehPsirelation} in the NS junction satisfies the corresponding boundary condition \eqref{boundaryconditioneh}. For example, an armchair boundary along the $x$-axis (with $M\propto\sigma_{x}$ independent of $x$) satisfies the requirement \eqref{MPrelation}, but a zigzag boundary along the $x$-axis ($M\propto\sigma_{z}$) does not. A pair of zigzag boundaries at $x=\pm L$ (with $M(\pm L,y)=\pm\sigma_{z}\otimes\nu_{z}$), on the other hand, do satisfy the requirement \eqref{MPrelation}.

The Dirac Hamiltonian \eqref{HDirac} of the \textit{p-n} junction may also contain an additional term $\delta H$ without spoiling the mapping to the NS junction, provided that $\delta H$ anticommutes with the operator product ${\cal PT}$,
\begin{equation}
{\cal PT}\delta H=-\delta H{\cal PT}.\label{deltaHPT}
\end{equation}
Considering the two examples of a $\delta H$ mentioned in Sec.\ \ref{TRS}, we see that the mass term preserves the mapping if $\mu(x,y)=\mu(-x,y)$, while the valley-dependent vector potential should satisfy ${\cal A}_{x}(-x,y)=-{\cal A}_{x}(x,y)$, ${\cal A}_{y}(-x,y)={\cal A}_{y}(x,y)$.

\subsection{\label{reflection} Retro-reflection vs.\ negative refraction}

\begin{figure}[tb]
\includegraphics[width=0.8\linewidth]{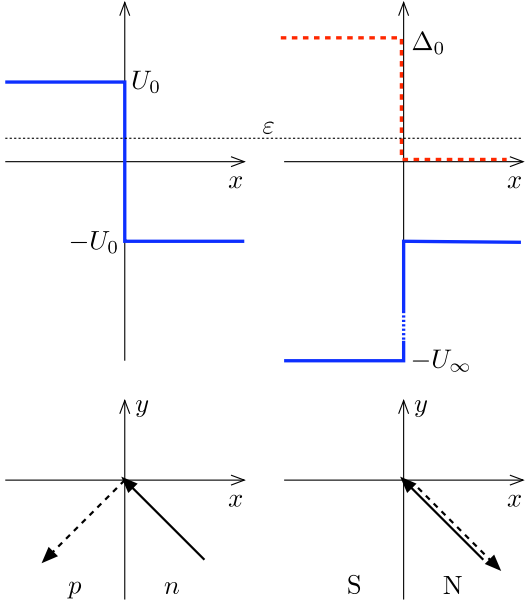}
\caption{\label{NSpn_reflection}
Comparison of two systems that can be mapped onto each other by the transformation \eqref{PsiehPsirelation}. The upper graphs show the electrostatic potential profile (solid lines) of a \textit{p-n} junction (left) and of the corresponding NS junction (right, with $U_{\infty}\gg U_{0}$). The upper right graph also shows the superconducting pair potential $\Delta$ (dashed line). The excitation spectrum of the two systems is the same for $\varepsilon\ll\Delta_{0}$. Classical trajectories in the two systems are related by reflection along the interface, as shown in the lower graphs for $\varepsilon=0$ (solid lines indicate the electron-like trajectories and dashed lines the hole-like trajectories).
}
\end{figure}

We apply the mapping of the previous subsection to an abrupt \textit{p-n} junction, as shown in Fig.\ \ref{NSpn_reflection}. With ``abrupt'' we mean that the width $d$ of the potential step at the \textit{p-n} interface should be small compared to the Fermi wave length $\lambda_{F}=hv/U_{0}$. In Sec.\ \ref{pn} we discussed the opposite regime $d\gg\lambda_{F}$ of a smooth interface, when only electrons approaching the interface near normal incidence are transmitted. For an abrupt interface the transmission probability is large also away from normal incidence, and an unusual effect of {\em negative refraction\/} appears \cite{Che07}: Upon crossing the \textit{p-n} interface the sign of the tangential velocity component is inverted. 

\begin{figure}[tb]
\includegraphics[width=0.6\linewidth]{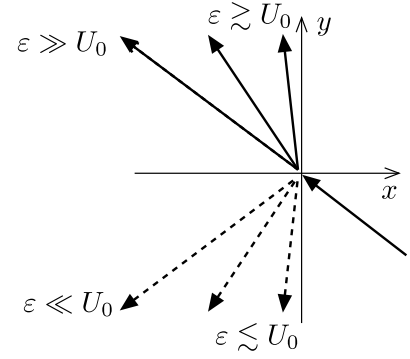}
\caption{\label{reflection_angles_eh}
Trajectories of an incident and refracted electron at a \textit{p-n} interface, for different excitation energies $\varepsilon$ relative to the potential step height $U_{0}$, at fixed angle of incidence. For $\varepsilon<U_{0}$ the refracted electron is in the valence band (dashed lines), while for $\varepsilon>U_{0}$ it is in the conduction band (solid lines). The refracted trajectories rotate counter-clockwise with increasing $\varepsilon$,  jumping by $180^{\circ}$ when $\varepsilon=U_{0}$. The transformation $x\mapsto -x$ maps this transition from negative to positive refraction onto the transition from retro-reflection to specular reflection in the NS junction of Fig.\ \ref{reflection_angles}.
}
\end{figure}

The lower panels in Fig.\ \ref{NSpn_reflection} show how the classical trajectories in the \textit{p-n} and NS junctions are mapped onto each other by reflection along the interface. Retro-reflection in the NS junction (inversion of the tangential velocity component upon conversion from electron to hole) maps onto negative refraction in the \textit{p-n} junction. As the excitation energy $\varepsilon$ increases beyond the step height $U_{0}$, negative refraction crosses over into positive refraction at the \textit{p-n} junction in the same way that retro-reflection crosses over into specular reflection at the NS junction (compare Figs.\ \ref{reflection_angles} and \ref{reflection_angles_eh}).

Because the mapping \eqref{PsiehPsirelation} is quantum mechanical, not only the trajectories are mapped onto each other but the full diffraction pattern together with the quantum mechanical transmission and reflection probabilities. For example, when $\varepsilon\ll U_{0},\Delta_{0}$ the NS junction has a probability $R_{eh}=\cos^{2}\theta$ for Andreev reflection (electron to hole) and a probability $R_{ee}=\sin^{2}\theta$ for normal reflection (electron to electron). This result \cite{Bee06} agrees with the transmission and reflection probabilities $T,R$ in an abrupt \textit{p-n} junction calculated by \textcite{Che06}, upon mapping $R_{eh}\mapsto T$ and $R_{ee}\mapsto R$. For normal incidence ($\theta=0$) both retro-reflection and negative refraction happen with unit probability. 

\begin{figure}[tb]
\includegraphics[width=0.8\linewidth]{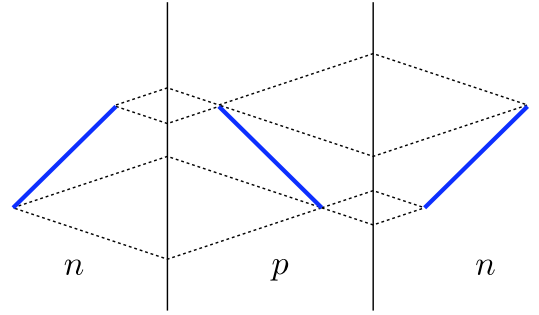}
\caption{\label{Veselago}
Classical trajectories (dotted lines) in an \textit{n-p-n} junction at an energy $\varepsilon=0$ that is halfway the potential step across the \textit{n-p} and \textit{p-n} interfaces, so that the refraction precisely inverts the angle of incidence. A scatterer in the \textit{n}-region (blue horizontal line) has an inverted image in the central \textit{p}-region and a noninverted image in the other \textit{n}-region. This is the principle of operation of the Veselago lens.
}
\end{figure}

Negative refraction was first discovered in optics,\footnote{
The most direct analogy is with the work of \textcite{Not00} on negative refraction in two-dimensional photonic crystals with the same honeycomb lattice as graphene.}
where it is used as a way to make a flat lens known as a Veselago lens \cite{Ves68}. [For a tutorial, see \textcite{Pen04}.] As calculated by \textcite{Che07}, an abrupt \textit{p-n} interface produces an inverted image in the \textit{p}-region of a scatterer in the \textit{n}-region. An \textit{n-p-n} or \textit{p-n-p} junction inverts the image twice, reproducing the original image at the other side of the junction (see Fig.\ \ref{Veselago}). The Veselago lens in graphene is not ideal: Negative refraction only produces a perfect focus at $\varepsilon=0$, while at other energies the focus is spread into a caustic. Caustics (focal lines, rather than focal points) also appear if the \textit{p-n} interface is curved rather than straight \cite{Cse07}.

\subsection{\label{QHE} Valley-isospin dependent quantum Hall effect}

In Sec.\ \ref{boundaries} we mentioned that the edge states in the lowest Landau level are valley polarized, with a valley isopin $\bm{\nu}$ determined by the boundary condition \eqref{boundarycondition} at the edge. Here we discuss how this valley polarization can be measured in a conduction experiment on either a \textit{p-n} junction or an NS junction.

\begin{figure}[tb]
\includegraphics[width=0.8\linewidth]{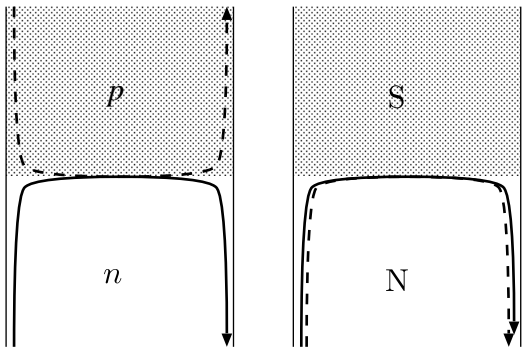}
\caption{\label{pnNS_analogy}
Schematic top view of a graphene nanoribbon containing an interface between a \textit{p}-doped and \textit{n}-doped region (left panel) and between a normal (N) and superconducting (S) region (right panel). Electron-like and hole-like edge states in the lowest Landau level are indicated by solid and dashed lines, respectively, with arrows pointing in the direction of propagation. \cite{Two07}
}
\end{figure}

The two geometries are compared in Fig.\ \ref{pnNS_analogy}. Electron-like and hole-like valley-polarized edge states hybridize along the \textit{p-n} or NS interface to form a valley-degenerate electron-hole state. (In the \textit{p-n} case, this state corresponds classically to the snake-shaped trajectory in Fig.\ \ref{NSpn_B}.) The two-terminal conductance $G=G_{0}T_{eh}$ is determined by the probability $T_{eh}$ that an electron-like state is converted into a hole-like state at the opposite edge (with $G_{0}=2e^{2}/h$ in the \textit{p-n} junction and $G_{0}=4e^{2}/h$ in the NS junction).\footnote{
One factor of two in $G_{0}$ comes from the spin degeneracy. The NS junction has one more factor of two because the electron-to-hole conversion transfers two electrons across the junction.}
As shown by \textcite{Akh07b} and \textcite{Two07}, in the absence of intervalley scattering this probability
\begin{equation}
T_{eh}=\tfrac{1}{2}(1-\cos\Phi) \label{Tehresult}
\end{equation}
depends only on the angle $\Phi$ between the valley isospins of the electron-like state at the two edges.

Eq.\ \eqref{Tehresult} assumes that the electron-like and hole-like edge channels at one edge have {\em opposite\/} valley isospins ($\pm\bm{\nu}_{L}$ for the left edge and $\pm\bm{\nu}_{R}$ for the right edge).\footnote{
This is generally the case, with one exception: A \textit{p-n} junction in a zigzag nanoribbon has electron-like and hole-like edge channels with {\em identical\/} valley isospins \cite{Two07}.} 
Since the unidirectional motion of the edge states prevents reflections, the total transmission matrix $t_{\rm total}=t_{R}t_{\rm int}t_{L}$ from one edge to the other edge is the product of  three $2\times 2$ unitary matrices: the transmission matrix $t_{L}$ from the left edge to the \textit{p-n} or NS interface, the transmission matrix $t_{\rm int}$ along the interface, and the transmission matrix $t_{R}$ from the interface to the right edge. In the absence of intervalley scattering $t_{\rm int}=e^{i\phi_{\rm int}}\tau_{0}$ is proportional to the unit matrix in the valley degree of freedom, while
\begin{equation}
t_{X}=e^{i\phi_{X}}|+\bm{\nu}_{X}\rangle\langle+\bm{\nu}_{X}|+e^{i\phi'_{X}}|-\bm{\nu}_{X}\rangle\langle-\bm{\nu}_{X}|\label{tpndef}
\end{equation}
(with $X=L,R$) is diagonal in the basis $|\pm\bm{\nu}_{X}\rangle$ of eigenstates of $\bm{\nu}_{X}\cdot\bm{\tau}$. The phase shifts $\phi_{\rm int},\phi_{X},\phi'_{X}$ need not be determined. Evaluation of the transmission probability 
\begin{equation}
T_{eh}=|\langle+\bm{\nu}_{L}|t_{\rm total}|-\bm{\nu}_{R}\rangle|^{2}\label{Tehdef}
\end{equation}
leads to the conductance
\begin{equation}
G=\tfrac{1}{2}G_{0}(1-\cos\Phi),\label{Gphi}
\end{equation}
with $\cos\Phi=\bm{\nu}_{L}\cdot\bm{\nu}_{R}$.

\begin{figure}[tb]
\includegraphics[width=1\linewidth]{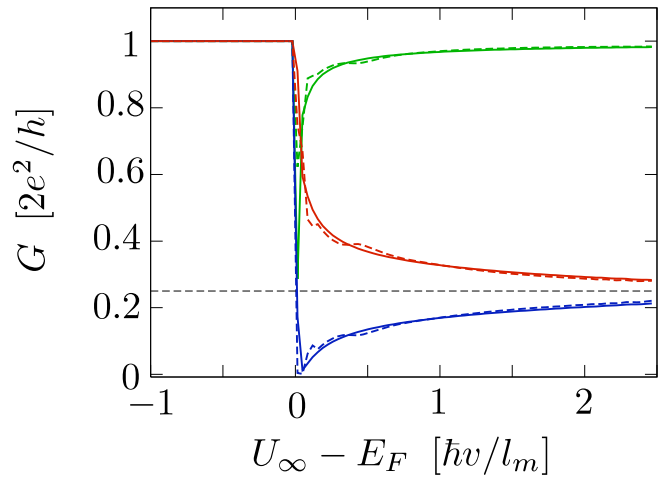}
\caption{\label{fig_disorder1}
Conductance of an armchair nanoribbon containing the potential step $U(x)=\frac{1}{2}[\tanh(2x/L)+1]U_{\infty}$, calculated numerically from a tight-binding model in a perpendicular magnetic field (magnetic length $l_{m}\equiv\sqrt{\hbar/eB}=5\,a$). The step height $U_{\infty}$ is varied from below $E_{F}$ (unipolar regime) to above $E_{F}$ (bipolar regime), at fixed $E_{F}=\hbar v/l_{m}$ and $L=50\,a$. The solid curves are without disorder, while the dashed curves are for a random electrostatic potential landscape (correlation length $\xi=10\,a$). The different colors correspond to a different number ${\cal N}$ of hexagons across the ribbon, and hence a different width $W=({\cal N}+3/2)a$: ${\cal N}=97$ (red curves), $98$ (blue), and $99$ (green). The dashed horizontal line marks the plateau at $G=\frac{1}{4}\times 2e^{2}/h$. \cite{Two07}
}
\end{figure}

The angle $\Phi=4\pi W/3a+\pi$ between the valley isospins at two opposite armchair edges depends on the width $W$ (as defined in Fig.\ \ref{monolayer}): $\Phi=\pi$ if $2W/a$ is a multiple of $3$, $|\Phi|=\pi/3$ if it is not (see Fig.\ \ref{Blochsphere}). A tight-binding model calculation of an armchair nanoribbon containing a potential step (Fig.\ \ref{fig_disorder1}) indeed shows that the conductance as a function of the step height switches from a plateau at the $\Phi$-independent Hall conductance $G_{0}$ in the unipolar regime (\textit{n-n} junction) to a $\Phi$-dependent value given by Eq.\ \eqref{Gphi} in the bipolar regime (\textit{p-n} junction). The plateau persists in the presence of a random potential, provided it is smooth on the scale of the lattice constant so no intervalley scattering is introduced.

The valley-isospin dependence of the quantum Hall effect makes it possible to use {\em strain\/} as a means of variation of the height of the conductance plateaus. As mentioned in Sec.\ \ref{TRS}, strain introduces a valley-dependent vector potential in the Dirac equation, corresponding to a fictitious magnetic field of opposite sign in the two valleys. This field rotates the Bloch vector of the valley isospin around the $z$-axis, which in the case of an armchair nanoribbon corresponds to a rotation of the valley isospin in the $x$-$y$ plane.

\begin{figure}[tb]
\includegraphics[width=0.8\linewidth]{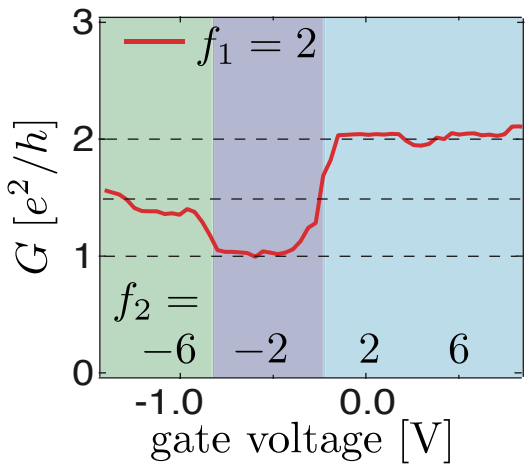}
\caption{\label{marcus}
Experimental conductance of a gate-controlled \textit{p-n} junction in graphene. The conductance of the \textit{n}-doped region at one side of the interface is fixed at $|f_{1}|e^{2}/h$, with $f_{1}=2$, while the conductance $|f_{2}|e^{2}/h$ at the other side of the interface is varied by the gate voltage (values of $f_{2}$ are indicated, with negative numbers corresponding to a \textit{p}-doped region). In the unipolar regime ($f_{1}f_{2}>0$) the conductance of the junction is given by $G=\min(|f_{1}|,|f_{2}|)e^{2}/h$, while in the bipolar regime ($f_{1}f_{2}<0$) the conductance is the Ohmic series conductance $G\times h/e^{2}=|f_{1}f_{2}|/(|f_{1}|+|f_{2}|)$. \cite{Wil07}
}
\end{figure}

In the high-magnetic field experiments of \textcite{Wil07} and \textcite{Ozy07} the \textit{p-n} junction has a quantized conductance, see Fig.\ \ref{marcus}. This has been explained by \textcite{Aba07} as the Ohmic series conductance $G_{\rm series}=G_{p}G_{n}/(G_{p}+G_{n})$ of the quantum Hall conductances $G_{p},G_{n}$ in the \textit{p}-doped and \textit{n}-doped regions (each an odd multiple of the conductance quantum $2e^{2}/h$). Ohm's law would apply if the system is sufficiently large that a local equilibrium is established at the interface, while the non-Ohmic result \eqref{Gphi} would be expected for smaller systems.

\subsection{\label{pseudo} Pseudo-superconductivity}

\begin{figure}[tb]
\includegraphics[width=1\linewidth]{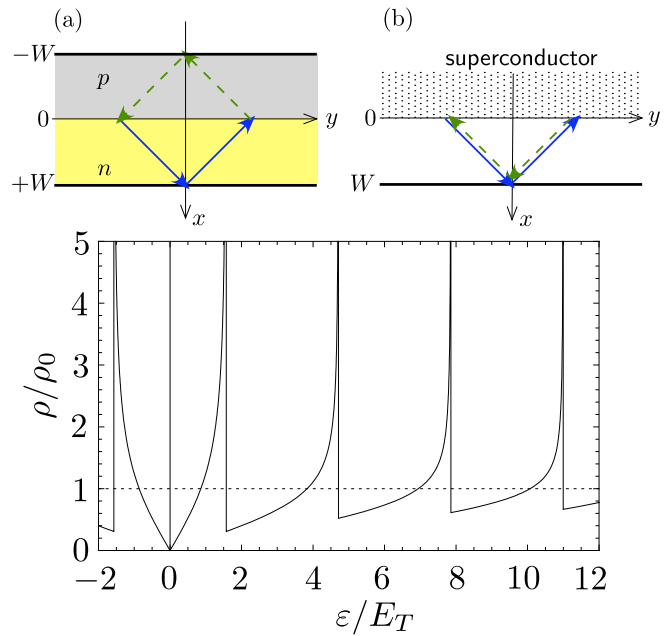}
\caption{\label{rhopn}
Plot of the density of states $\rho(\varepsilon)$ for the \textit{p-n} junction shown in (a). The dotted line is the value in the isolated \textit{p} and \textit{n} regions, which is energy independent for $|\varepsilon|\ll U_{0}$. The density of states vanishes at the Fermi level ($\varepsilon=0$), according to Eq.\ \eqref{rhoresult}. The NS junction shown in (b) has the same density of states. In both the NS and \textit{p-n} geometries the suppression of the density of states is due to destructive interference of the electron-like and hole-like segments of the periodic orbits at the Fermi level (indicated by solid blue and dashed green trajectories). \cite{Bee07}
}
\end{figure}

The correspondence between NS and \textit{p-n} junctions of Sec.\ \ref{mapping} implies that quantum effects associated with superconductivity, such as the proximity effect and the Josephson effect, have analogues in nonsuperconducting bipolar graphene \cite{Bee07}.

Such ``pseudo-superconductivity'' is demonstrated in Fig.\ \ref{rhopn}, which plots the density of states $\rho(\varepsilon)$ in a \textit{p-n} junction with an abrupt interface. The \textit{p} and \textit{n} regions have the same Fermi energy $U_{0}$ and zigzag boundaries at $x=\pm W$. The width $W$ is assumed to be large compared to the Fermi wave length $\lambda_{F}=hv/U_{0}$. The density of states, smoothed over rapid oscillations, vanishes linearly as
\begin{equation}
\rho(\varepsilon)=\rho_{0}|\varepsilon|/E_{T}\label{rhoresult}
\end{equation}
for small $|\varepsilon|$, with $\rho_{0}=(2U_{0}/\pi)(\hbar v)^{-2}$ the density of states (per unit area and including spin plus valley degeneracies) in the separate \textit{p} and \textit{n} regions. The energy $E_{T}=\hbar v/2W$ is the Thouless energy (which is $\ll\mu$ for $W\gg\lambda_{F}$). This suppression of the density of states at the Fermi level by a factor $\varepsilon/E_{T}$ is precisely analogous to an NS junction, where the density of states is suppressed by the superconducting proximity effect \cite{Tit07}. In particular, the peaks in $\rho(\varepsilon)$ at $\varepsilon=\pi E_{T}(n+1/2)$, $n=0,1,2,\ldots$, are analogous to the De Gennes-Saint James resonances in Josephson junctions \cite{DeG63}.

In a semiclassical description, the suppression of the density of states in the \textit{p-n} junction can be understood as destructive interference of the electron-like and hole-like segments of a periodic orbit (solid and dashed lines in Fig.\ \ref{rhopn}a). At the Fermi level, the dynamical phase shift accumulated in the \textit{p} and \textit{n} regions cancels, and what remains is a Berry phase shift of $\pi$ from the rotation of the pseudospin of a Dirac fermion.

\begin{figure}[tb]
\includegraphics[width=.9\linewidth]{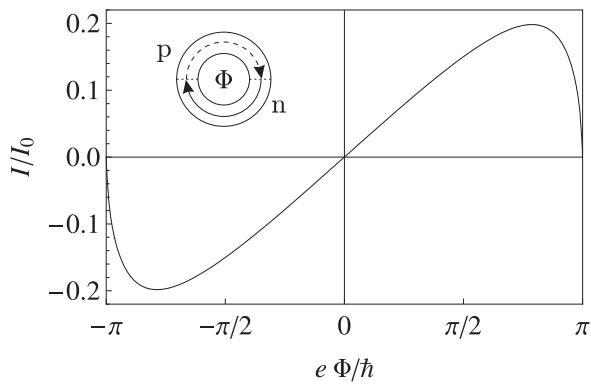}
\caption{\label{current}
Persistent current through a ring containing an abrupt \textit{p-n} interface, as a function of the magnetic flux through the ring. \cite{Bee07}
}
\end{figure}

If the \textit{p} and \textit{n} regions enclose a magnetic flux $\Phi$, as in the ring geometry of Fig.\ \ref{current} (inset), then the Berry phase shift can be compensated and the suppression of the density of states can be eliminated. The resulting flux dependence of the ground state energy $E={\cal A}\int_{-\infty}^{0}\rho(\varepsilon)\varepsilon\,d\varepsilon$ (with ${\cal A}$ the joint area of the \textit{n} and \textit{p} regions) implies that a current $I=dE/d\Phi$ will flow through the ring in equilibrium, as in a Josephson junction \cite{Imr97}. According to Eq.\ \eqref{rhoresult}, the order of magnitude
\begin{equation}
I_{0}=(e/\hbar)E_{T}^{2}/\delta=(e/\hbar)NE_{T}\label{I0def}
\end{equation}
of this persistent current is set by the level spacing $\delta=({\cal A}\rho_{0})^{-1}$ and by the Thouless energy $E_{T}=\hbar v/\pi r=N\delta$ in the ring geometry (of radius $r$ and width $w\ll r$, supporting $N=4U_{0} w/\pi\hbar v\gg 1$ propagating modes). Because of the macroscopic suppression of the density of states, this is a macroscopic current --- larger by a factor $N$ than the mesoscopic persistent current in a ballistic metal ring \cite{Imr97,But83}.

Fig.\ \ref{current} plots $I(\Phi)$ for an abrupt \textit{p-n} junction in an $N$-mode ring without intermode scattering \cite{Bee07}. The maximal persistent current is $I_{c}\approx 0.2\,I_{0}$. Up to a numerical coefficient, this result for $I_{c}$ is the same as the critical current of a ballistic Josephson junction.\footnote{
For a detailed comparison of the persistent current through the bipolar junction and the supercurrent through the analogous Josephson junction, see the Appendix of arXiv:0710.1309.
}

This concludes our review of Andreev reflection and Klein tunneling in graphene. The analogies discussed in this Section will hopefully be validated in the near future by ongoing experiments on bipolar junctions and Josephson junctions. From a different perspective, the correspondence between these two phenomena offers the intriguing opportunity to observe superconducting analogies in non-electronic systems governed by the same Dirac equation as graphene. An example would be a two-dimensional photonic crystal on a honeycomb or triangular lattice \cite{Rag06,Sep07,Gar07}.

\acknowledgments
The research from my own group reported in this Colloquium was performed in collaboration with A. R. Akhmerov, A. Ossipov, P. Recher, A. Rycerz, I. Snyman, M. Titov, B. Trauzettel, and J. Tworzyd{\l}o. It was supported by the Dutch Science Foundation NWO/FOM.

\bibliographystyle{apsrmp}

\end{document}